\documentclass[prb,twocolumn,aps]{revtex4-2}
\usepackage{graphicx}
\usepackage{amsmath,amssymb,dsfont,bbold}
\usepackage{hyperref}

\begin{document}
\title{Pairing and Pair Breaking by Gauge Fluctuations in Bilayer Composite Fermion Metals}

\author{Haoyun Deng} 
\author{Luis Mendoza}
\altaffiliation{Current Address: Fermi National Accelerator Laboratory, P.O. Box 500, Batavia, IL 60510, USA}
\author{N.E.~Bonesteel}
\affiliation{Department of Physics and National High Magnetic
Field Laboratory, Florida State University, Tallahassee, FL 32310,
USA}

\date{\today}

\begin{abstract}
We study interlayer pairing of composite fermions in the total $\nu=1/2+1/2$ quantum Hall bilayer as a possible framework for understanding the experimentally observed transition from a compressible state at large layer spacing to a bilayer quantum Hall state at small layer spacing.  We consider a model in which the effective interlayer composite fermion pairing interaction mediated by the Chern-Simons gauge fields in the two layers is singular with both attractive (out-of-phase) and repulsive (in-phase) components diverging at low frequency. If only the more singular attractive interaction is included the pairing gap obtained by solving the gap equation is proportional to the inverse of the layer spacing squared.  In the so-called local approximation, we find that when the less singular repulsive interactions are also included the pairing gap still falls off as inverse layer spacing squared, consistent with recent analyses, but is strongly suppressed to a degree that may account for the fact that this predicted inverse square dependence is not observed experimentally. The analytically obtained local approximation solutions are then used as a starting point to numerically iterate the full gap equation to assess the validity of the approximation in this limit. 
\end{abstract}

\maketitle

\section{Introduction}

The total $\nu = 1$ quantum Hall bilayer consists of two parallel two-dimensional electron gases separated by a distance $d$ and placed in a perpendicular magnetic field $B$ such that the total electron density of the two layers is that of a single filled Landau level.  For the symmetrically doped case, each layer has Landau level filling fraction $\nu=1/2$, and, if interlayer electron tunneling can be ignored, the only coupling between layers is through the interlayer Coulomb repulsion.  The strength of this interlayer repulsion, relative to the intralayer Coulomb repulsion, is set by the dimensionless ratio $d/l_0$, where $l_0 = (\hbar c/(eB))^{1/2}$ is the magnetic length. 

In the limit of small $d/l_0$ this system forms an incompressible bilayer quantum Hall state in which electrons develop spontaneous interlayer phase coherence \cite{moon95}.   This state can be viewed as an exciton condensate formed by electron-hole pairs in the two layers \cite{dassarma_book97,eisenstein04}.  In the absence of interlayer tunneling, the transition to this state is predicted to occur at finite temperature via a Kosterlitz-Thouless transition.  In the opposite limit of large $d/l_0$, the two layers decouple and the system is presumably well described in terms of two separate $\nu=1/2$ composite fermion metals. In these compressible states, composite fermions, electrons attached to two fictitious (Chern-Simons) flux quanta \cite{jain89_prl,jain89_prb,jainbook,halperin93}, move in zero effective magnetic field and are predicted to experience a fluctuating effective gauge field.  While there has been some speculation that the correct low-energy description of the resulting compressible state is in terms of Dirac particles \cite{son15}, we take the view here that even if this is the case, the effective Halperin, Lee, Read description is still essentially correct, provided it is applied appropriately.

It has been argued recently that the appearance of the electron-hole exciton condensate in this system may be equivalent to the formation of an interlayer $p$-wave composite fermion superconductor \cite{sodemann17}, a type of interlayer pairing of composite fermions first studied in \cite{bonesteel96}.  This viewpoint has gained experimental support in the recent work of \cite{liu22} in which quantum Hall states in graphene double layers separated by atomically thin insulators were studied.  In this work, through measurement of the Coulomb drag responses (both longitudinal and transverse drag resistivities) and counterflow responses of the graphene layers at total filling fraction $\nu = 1$ as a function of temperature, a temperature vs. $d/l_0$ phase diagram was determined.  This phase diagram showed that in the limit of small $d/l_0$, there is indeed a finite temperature phase transition into a quantum Hall state, and for temperatures above this transition there is a wide temperature range in which interlayer pairs have formed (as evidenced by the counterflow response), but the system does not have true long-range order and hence does not show the quantum Hall effect (as observed in the drag response).  This behavior is characteristic of a crossover from a regime in which tightly bound pairs form above the Bose-Einstein condensation temperature for these pairs, to a regime, with increasing $d/l_0$ in which the pairs appear to form at temperatures just above the critical temperature, as expected for more conventional weak coupling BCS pairing, and so is at least consistent with the notion of interlayer pairing.

A full random-phase approximation (RPA) calculation of the effective interlayer pairing interaction for composite fermions in this system \cite{isobe17} shows that within this approximation the leading pairing instability is in the $p$-wave channel, and so may correspond to the experimentally observed quantum Hall state as suggested in \cite{sodemann17}.  More recently this result has also been obtained using a modified RPA which accounts for the renormalization of the composite fermion mass required by the lowest Landau level projection \cite{lotrivc23}.  This latter work also provides evidence that this composite fermion pairing approach is equivalent to a third viewpoint (in addition to the exciton condensate view) in which particle-hole symmetry of the half-filled Landau level is exploited and the bilayer system is viewed in terms of pairing of ``composite electrons" in one layer with ``composite holes" in the second layer \cite{liu22,ruegg23}. All this, along with mounting numerical evidence \cite{moeller08,moeller09,wagner21} that suggests these approaches are all equivalent and that interlayer pairing of composite fermions of the type originally studied in \cite{bonesteel96} may indeed be an appropriate framework for studying the transition from bilayer composite fermion metal to bilayer quantum Hall state, motivates the present work.  

In this paper, we analyze the effect of both pairing and pair-breaking gauge fluctuations on bilayer pairing as a function of layer spacing.  We focus on the balanced case and take the point of view that the system can be described for large layer spacing as two composite fermion metals whose low-energy physics is described by the Halperin-Lee-Read theory, with appropriately renormalized effective mass.  We keep only the most singular interactions in the interlayer Cooper channel which have the same form in all angular momentum channels, and treat the less singular terms, which presumably determine the pairing symmetry of the gap, phenomenologically. Here we are not focused on determining the pairing channel with the leading instability, something which is addressed by more detailed RPA calculations \cite{isobe17, ruegg23, lotrivc23}, but rather on the question of how strong the pairing is, to what extent gauge fluctuation pairing might play a role in it, and how it depends on layer spacing when both pairing and pair breaking contributions are taken into account.  The primary tool we use for our analysis is the so-called local approximation applied to the frequency-dependent $T=0$ gap equation.  Within this approximation, we derive analytic results for the combined effect of pairing and pair breaking in this system.

The paper is organized as follows.  In Section \ref{gauge} we review the Halperin, Lee, Read description of the the $\nu =1/2+1/2$ bilayer and present our model frequency-dependent interaction.  Section \ref{gap_eq} then presents a simple example of the local approximation applied to the $T=0$ gap equation due to both the singular pairing and pair-breaking gauge fluctuations.  In Section \ref{lls} we focus on the large layer spacing limit, obtaining analytic expressions for the $T=0$ gap, as well as the significance of pair-breaking effects.  We find that the latter are characterized by a single ``pair breaking" parameter and are actually quite large for the parameters relevant to the bilayer system.   Section \ref{xover} then describes our results for what happens as the short-range, nonsingular interaction strength is varied, and the crossover from the weak gauge pairing we associate with the large layer spacing limit, to the strong BCS pairing due to short-range attraction in some angular momentum channel.   In Sections \ref{lls} and \ref{xover} we also compare the results of the local approximation to exact solutions of the gap equation obtained numerically, verifying the essential validity of the approximation. Finally, in Section \ref{conc} we review the results of the paper and present our conclusions.

\section{Pairing and Pair-breaking Gauge Fluctuations}
\label{gauge}

In what follows we assume that the Halperin, Lee, Read description of the $\nu=1/2$ single layer state adequately captures the essential low-energy physics of the system.  In this approach, the system is described with a Euclidean-time action for the $\nu=1/2+1/2$ bilayer which, at temperature $T$, has the form $S = \int_0^\beta d \tau \int d^2 r {\cal L}({\bf r},\tau)$ where $\beta = (k_B T)^{-1}$ and the Lagrangian density is ${\cal L} = {\cal L}_0 + {\cal L}_{int} + {\cal L}_{CS}$ with
\begin{eqnarray}
{\cal L}_0 = \sum_{\alpha=\uparrow,\downarrow} \overline{\psi}_\alpha \Bigl(\partial_\tau - i a_0^\alpha   - \frac{1}{2m^*}(\nabla
- i {\bf a}^\alpha)^2 \Bigr) \psi_\alpha,
\label{L0}
\end{eqnarray}
\begin{equation}
{\cal L}_{int} = \frac12 \sum_{\alpha,\alpha^\prime = \uparrow,\downarrow} \int d^2 r^\prime~ \delta \rho_\alpha({\bf r},\tau)V_{\alpha\alpha^\prime}({\bf r} - {\bf r}^\prime) \delta \rho_\alpha^\prime({\bf r}^\prime,\tau), \label{Lint}
\end{equation}
where $\uparrow,\downarrow$ are pseudospin labels for the layers,
\begin{eqnarray}
V_{\uparrow \uparrow}({\bf r}) = V_{\downarrow\downarrow}({\bf r}) &=& \frac{e^2}{\epsilon r},\\
V_{\uparrow\downarrow}({\bf r}) = V_{\downarrow\uparrow}({\bf r}) &=& \frac{e^2}{\epsilon \sqrt{r^2 + d^2}},
\end{eqnarray}
are the interlayer and intralayer Coulomb repulsion, respectively, and
\begin{eqnarray}
{\cal L}_{CS} = -\frac{i}{2\pi\lambda} \sum_{\alpha = \uparrow,\downarrow} a_0^\alpha {\hat {\bf z}} \cdot \left(\nabla \times \left({\bf a}^\alpha + e{\bf A}_{ext}\right)\right),
\label{LCS}
\end{eqnarray}
is the Chern-Simons action in the Coulomb gauge.  Here $\psi_\alpha$ is the composite fermion field in layer $\alpha = \uparrow,\downarrow$, ${\bf A}_{ext}$ is the vector potential for the external applied magnetic field ${\bf B} = \nabla \times {\bf A}_{ext} = {\bf \hat z} 2\pi\tilde\phi n/e$ where $n$ is the electron density in each layer, $\tilde{\phi}$ is the number of flux quanta attached to each composite fermion ($\tilde{\phi} = 2$ for the case total $\nu=1 = 1/2+1/2$ considered here), $m^*$ is the effective mass of the composite fermions, and $(a_0^\alpha, {\bf a}^\alpha)$ is the Chern-Simons gauge field seen by composite fermions in layer $\alpha$.  ${\cal L}_{CS}$ is a Chern-Simons term in the Coulomb gauge for which $\nabla \cdot {\bf a}^\alpha = 0$.   The gauge field degrees of freedom in each layer then consist entirely of the time component, $a^\alpha_0$, and transverse component, $a^\alpha_1({\bf q},\tau) = {\bf {\hat z}} \cdot ({\bf {\hat q}} \times {\bf a}^\alpha ({\bf q}, \tau))$, of the Chern-Simons gauge fields.  

The partition function for this system is then ${\cal Z} = \int \prod_{\alpha=\uparrow,\downarrow} D\psi_\alpha D a^\alpha_0 D a^\alpha_1 e^{-S}$. Integrating out $a^\alpha_0$ enforces the constraint
\begin{eqnarray}
\delta \rho_\alpha = \frac{1}{2\pi\tilde{\phi}} {\bf \hat z} \cdot \nabla \times {\bf a}^\alpha, \label{flux_attach}
\end{eqnarray}
which describes attaching $\tilde\phi$ quanta of Chern-Simons flux to the composite fermions.  Using (\ref{flux_attach}) the density fluctuations in (\ref{Lint}) can be replaced with transverse gauge fields and the resulting action is purely quadratic in the fermions. The RPA is then carried out by integrating out these fermions and expanding the resulting effective action to second order in the gauge fields. The result of this procedure is an effective action $S_{RPA} = S^+_{RPA} + S^-_{RPA}$, where
\begin{equation}
S^\pm_{RPA}  =\frac{1}{2}\sum_{{\omega_n, {\bf q}\atop{\mu,\nu=0,1}}} {a^\pm_\mu}^*({\bf q},i\omega_n) {{{\cal D}^\pm}}^{-1}_{\mu\nu} ({\bf q},i\omega_n;\Phi) a^\pm_\nu({\bf q},i\omega_n). 
\label{RPA_action}
\end{equation}
Here
\begin{eqnarray}
a^+_\mu &=& (a^\uparrow_\mu + a^\downarrow_\mu)/\sqrt{2}, \label{ap}\\
a^-_\mu &=& (a^\uparrow_\mu - a^\downarrow_\mu)/\sqrt{2}, \label{am}
\end{eqnarray}
where $\omega_n$ are bosonic Matsubara frequencies, $a^\pm_0({\bf q},i\omega_n)$ and $a^\pm_1({\bf q},i\omega_n) = {\bf{\hat z}} \cdot ({\bf \hat q} \times {\bf a}^\pm({\bf q},i\omega_n))$ are, respectively, the time and transverse components of the gauge fields, and ${{\cal D}^\pm}_{\mu\nu}^{-1}$ is the inverse of the gauge propagator matrix.

The result for the gauge field propagators ${\cal D}_{\mu\nu}^\pm(q,\omega)$ can be computed analytically for all $q$ and $\omega$ within the RPA \cite{cipri14, isobe17}.  Here, as in \cite{bonesteel93,bonesteel96}, we focus on the low-energy long wavelength fluctuations.  In this limit, the dominant interactions are mediated by the in-phase and out-of-phase transverse gauge fluctuations \cite{bonesteel93}.  For $\tilde \phi = 2$ and in the large layer spacing limit the result is,
\begin{eqnarray}
{\cal D}^-_{11}(q,\omega) &\simeq& \frac{1}{\frac{e^2}{4\pi\epsilon} d q^2 + \frac{1}{4\pi l_0} \frac{|\omega|}{q}}, \\
{\cal D}^+_{11}(q,\omega) &\simeq& \frac{1}{\frac{e^2}{4\pi\epsilon} q + \frac{1}{4\pi l_0} \frac{|\omega|}{q}}, 
\end{eqnarray}
The out-of-phase gauge fluctuations are seen to be more singular than the in-phase gauge fluctuations. This is a consequence of the fact that the transverse gauge field fluctuations are directly tied to density fluctuations, and out-of-phase density fluctuations are stronger than in-phase density fluctuations due to the interlayer Coulomb repulsion.

In addition to being more singular, the out-of-phase fluctuations mediate a strong attractive interaction in the interlayer Cooper channel for composite fermions.  This is because the out-of-phase gauge fields couple to composite fermions in the two layers as if they had opposite charge.  The in-phase fluctuations, on the other hand, while less singular, are repulsive in the Cooper channel and so suppress pairing. 

Here we are concerned with the net interlayer pairing mediated by these gauge fluctuations, and in particular the interplay between the singular pairing and pair-breaking terms.  To analyze this we use the frequency-dependent couplings associated with the most singular gauge field contributions described above \cite{bonesteel93,wang14},
\begin{eqnarray}
\lambda_\pm(\omega) =  \left\langle \left|\frac{{\bf k} \times \hat{\bf k}^\prime}{m^*}\right|^2 D^\pm_{11}(|{\bf k} - {\bf k}^\prime|,\omega) e^{il\theta_{{\bf k},{\bf k^\prime}}}\right\rangle_{{\bf k},{\bf k}^\prime \in {\rm FS}} \label{FS_average}
\end{eqnarray}
where the angle brackets indicate a Fermi surface average over ${\bf k}$ and ${\bf k}^\prime$. 
Note that here, and all that follows, we work at $T=0$ and have analytically continued to imaginary frequency.  In the low-energy long wavelength limit, and in the limit of large layer spacing, one finds (l.s.t. = less singular terms),
\begin{eqnarray}
\lambda_-(\omega) &=& \frac{8}{3\sqrt{3}} \frac{1}{\beta^{2/3}}\left(\frac{l_0}{d}\right)^{2/3} \left|\frac{\omega_0}{\omega}\right|^{1/3}
 + {\rm l.s.t.},\\
\lambda_+(\omega) &=& \frac{2}{\pi} \frac{1}{\beta} \ln \left|\frac{\omega_0}{\omega}\right|
+ {\rm l.s.t.},
\end{eqnarray} 
where $\omega_0 = e^2/(\epsilon l_0)$ sets the typical Coulomb energy scale of the problem within the lowest Landau level, and $\beta = e^2 l_0 m^*/\epsilon$ is a dimensionless parameter which we take to be of order 1 to account for the required renormalization of the composite fermion mass to a scale set by the Coulomb repulsion, with $m^* \sim \epsilon/(e^2 l_0)$.  

Here the integer $l$ appearing in the $e^{il\theta_{{\bf k},{\bf k}^\prime}}$ factor in (\ref{FS_average}), where $\theta_{{\bf k},{\bf k}^\prime}$ is the angle between the vectors ${\bf k}$ and ${\bf k}^\prime$, is the angular momentum pairing channel.  The leading singular contributions to $\lambda_+$ and $\lambda_-$ are independent of $l$ due to the fact that the gauge field mediated interactions are singular at small $q$.  All the $l$ dependence in the couplings appears in the less singular terms. 

\section{Gap Equation and Local Approximation}
\label{gap_eq}

The simplest approach to the essential physics of this bilayer pairing is to solve the $T=0$ frequency-dependent gap equation for this problem (again, here $\omega$ and $\Omega$ are imaginary frequencies),
\begin{eqnarray}
\Delta(\omega) &=& \frac{1}{2} \int d\Omega \lambda(\omega - \Omega) \frac{\Delta(\Omega)}{\sqrt{\Omega^2 + |\Delta(\Omega)|^2}}.\label{gap_equation}
\end{eqnarray}
Here $\lambda(\omega)$ is the frequency-dependent effective interaction, including contributions described in Section \ref{gauge}.  Precisely this problem was studied numerically in \cite{wang14}.  Here we focus on understanding the nature of the pairing and the interplay between the pairing and pair-breaking gauge fluctuations.  To this end, it is useful to have approximate analytic solutions, which can provide some physical insight into the behavior of the system, in addition to exact numerical solutions. 

Specifically, we are interested in solving (\ref{gap_equation}) for the case
\begin{eqnarray}
\lambda(\omega) &=& (\lambda_-(\omega) - \lambda_+(\omega) - V_0) \Theta(|\omega| - \Lambda),\label{model}
\end{eqnarray}
where 
\begin{eqnarray}
\lambda_-(\omega) = \alpha_- \Bigl | \frac{\omega_0}{\omega} \Bigr |^\gamma,\\
\lambda_+(\omega) = \alpha_+ \log \Bigl|\frac{\omega_0}{\omega}\Bigr|.
\end{eqnarray}
We take the cutoff $\Lambda$ to be $\omega_0$ and view the parameters $\alpha_-$, $\alpha_+$ and $V_0$ as effective parameters.   From the RPA result described in Sec.~\ref{gauge} we see that for the bilayer we have $\gamma = 1/3$, and in the large layer spacing limit $\alpha_- \simeq (l_0/d)^{2/3}$, $\alpha_+ \simeq 1$, and $\omega_0 \simeq e^2/(\epsilon l_0)$ \cite{bonesteel93}.  Here $V_0$ is a nonsingular term that represents the short-range interaction and which will depend on the angular momentum channel.  We make no attempt to calculate $V_0$ from first principles here, as in \cite{isobe17} and \cite{lotrivc23}, but rather treat it as a semi-phenomenological parameter.  The rest of the paper focuses on finding approximate and exact solutions of (\ref{gap_equation}) for the interaction (\ref{model}) and determining how these solutions depend on $\alpha_-, \alpha_+,$ and $V_0$.

We now apply the local approximation, as described in \cite{wang17}, to this pairing problem.  The first step in this approximation is to linearize the gap equation,
\begin{eqnarray}
\Delta(\omega) &=& \frac{1}{2} \int_{|\Omega| > |\Delta(0)|} d\Omega \lambda(\omega - \Omega) \frac{\Delta(\Omega)}{|\Omega|}.
\end{eqnarray}
Taylor expanding $\lambda(\Omega-\omega)$ inside the integral in the following way,
\begin{eqnarray}
\lambda(\Omega -\omega) = \left\{\begin{array}{cc}
\lambda(\omega) - \Omega \lambda^\prime(\omega) + \cdots,&|\Omega| < |\omega| \\
\lambda(\Omega) -\omega \lambda^\prime(\Omega) + \cdots, & |\Omega| > |\omega|\end{array}\right.
\label{localapp}
\end{eqnarray}
and keeping just the first term, we obtain a linear integral equation, which upon successive derivatives with respect to $\omega$ can be shown to be mathematically equivalent to the following second-order linear differential equation,
\begin{eqnarray}
\frac{d}{d\omega} \left(\frac{\Delta^\prime(\omega)}{\lambda(\omega)}\right) - \frac{\Delta(\omega)}{\omega} = 0,
\end{eqnarray}
with boundary conditions
\begin{eqnarray}
\Delta^\prime(\omega) \Bigl|_{\omega = \Delta_0} &=& 0,
\end{eqnarray}
and
\begin{eqnarray}
\frac{d}{d\omega}\left(\frac{\Delta(\omega)}{\lambda(\omega)}\right)\Bigl|_{\omega = \Lambda} = 0.
\end{eqnarray}

This second-order linear differential equation can be converted into a nonlinear first-order differential equation by introducing the $V$ function,
\begin{eqnarray}
V(\omega) = - \lambda^\prime(\omega) \frac{\Delta(\omega)}{\Delta^\prime(\omega)}.
\end{eqnarray}
If we further introduce a flow parameter,
\begin{eqnarray}
l = \ln \frac{\Lambda}{\omega},
\end{eqnarray}
the differential equation can be converted to the following,
\begin{eqnarray}
\frac{dV}{dl} = -\frac{d\lambda}{dl} - V^2.
\end{eqnarray}
The UV boundary condition becomes,
\begin{eqnarray}
V(\Lambda) = -\lambda(\Lambda),
\end{eqnarray}
and the IR boundary condition becomes,
\begin{eqnarray}
\lim_{\omega \rightarrow \Delta(0)} V(\omega) = -\infty,
\end{eqnarray}
where here $\Delta(0)$ is the zero frequency gap.

\begin{figure}[t]
  \begin{center}
	 \includegraphics[width=2.7in]{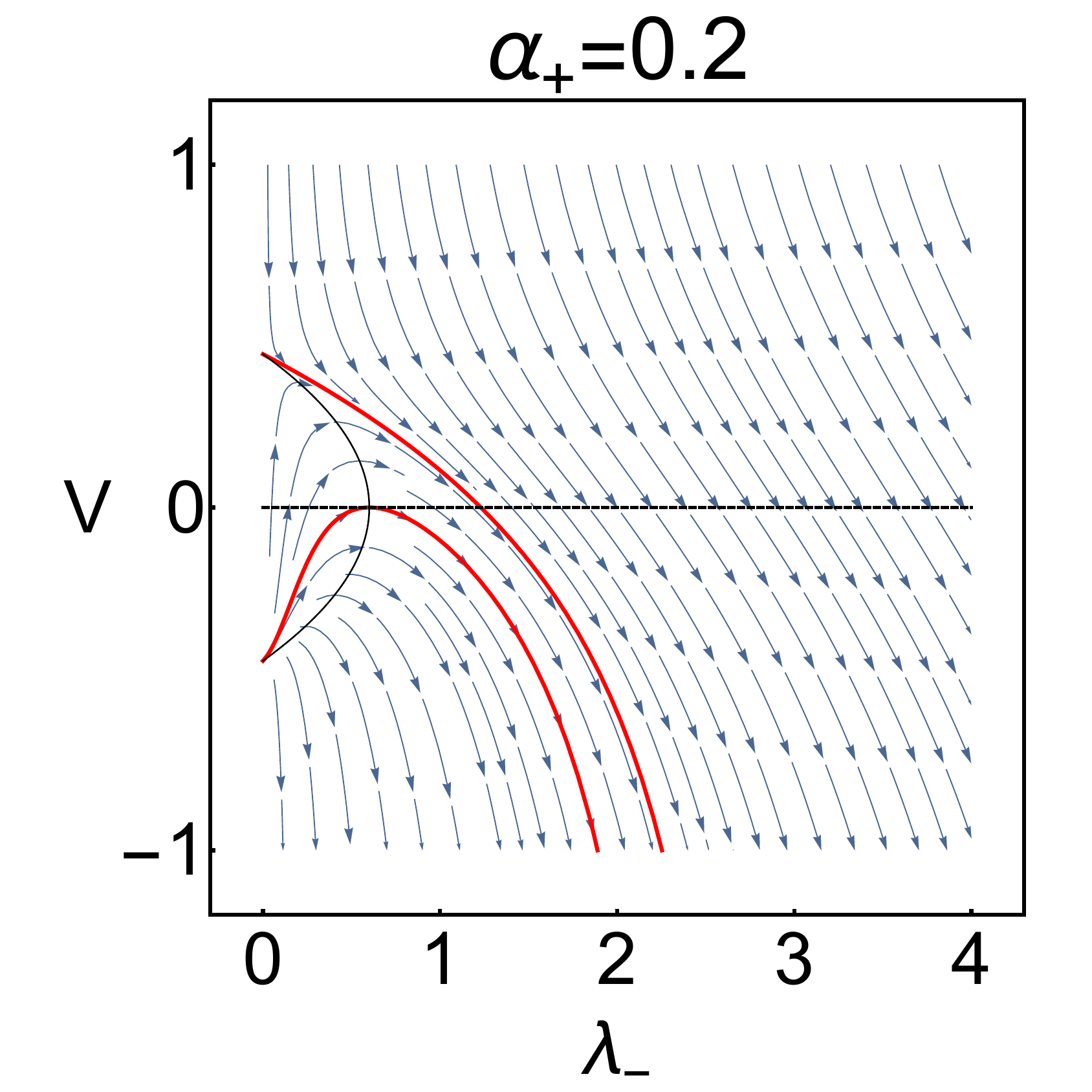}\\
	\includegraphics[width=2.7in]{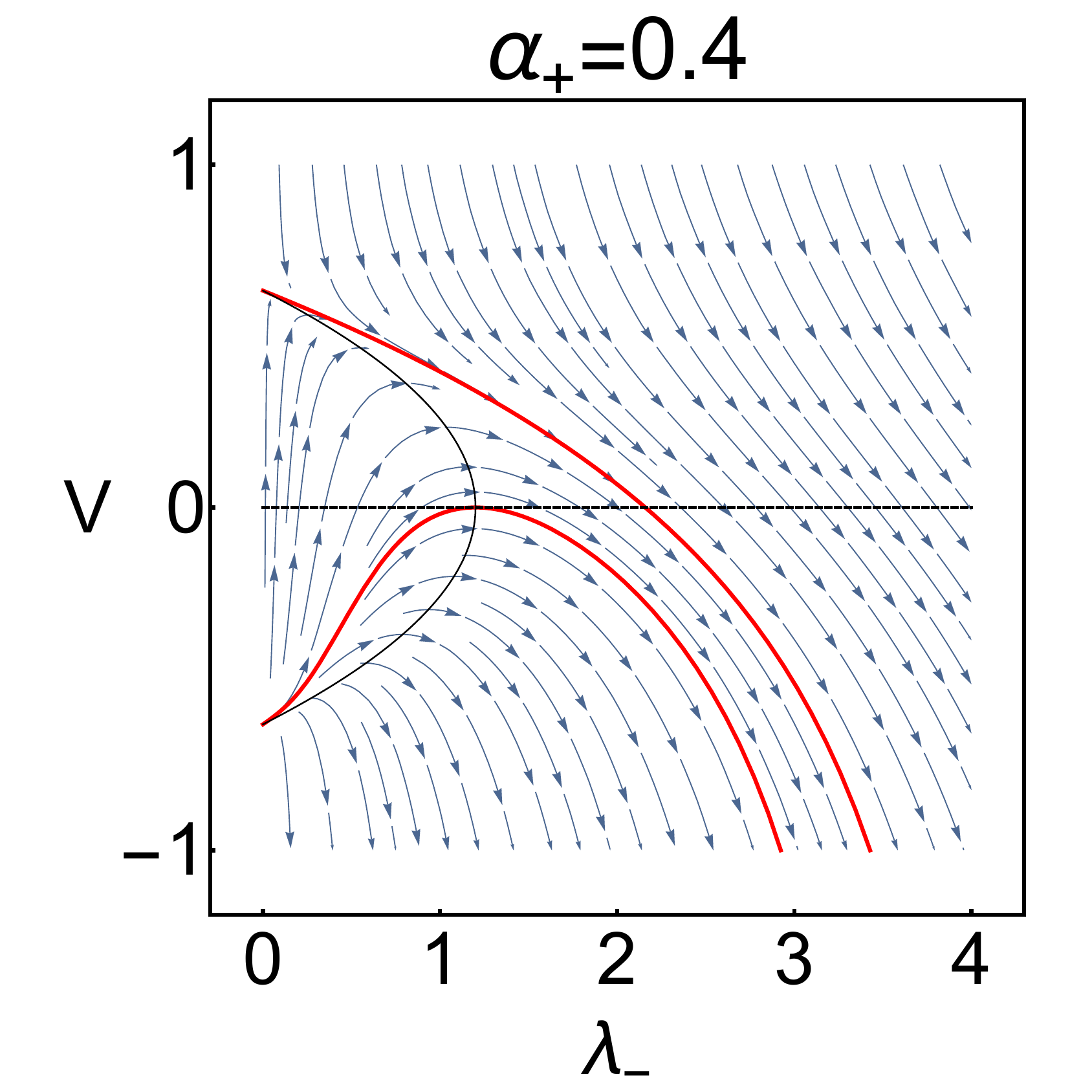}	 
  \caption{(Color online)  $V, \lambda_-$ flow diagrams for the case $\alpha_+ = 0.2$ and $\alpha_+ = 0.4$. In each diagram the top red trajectory corresponds to the large layer spacing limit solution, the lower red trajectory corresponds to the crossover from gauge pairing to BCS pairing, as defined in the text, and the solid black curves connect $V = +\sqrt{\alpha_+}$ with $V = - \sqrt{\alpha_+}$ at $\lambda_- = 0$, following the curve $V^2 = \alpha_+ - \gamma \lambda_-$, along which $\frac{dV}{dl} = 0$.}\label{flow}
   \end{center}
\end{figure}

The frequency-dependent gap function $\Delta(\omega)$ is then simply given by the following,
\begin{eqnarray}
\ln \Delta = - \int \frac{\lambda^\prime}{V} d\omega, \label{vtod}
\end{eqnarray}
where here we keep the right-hand side as an indefinite integral because the integration constant is just a multiplicative constant in $\Delta$.  

For the case $\lambda(\omega) = \lambda_-(\omega) - \alpha_+ \log \left|\frac{\omega}{\omega_0}\right| - V_0$ the flow equation can be written
\begin{eqnarray}
\frac{dV}{dl} = -\gamma \lambda_- + \alpha_+ - V^2. \label{flow1}
\end{eqnarray}
This equation is analogous to the renormalization group equations given in \cite{sodemann17} where the non-Fermi liquid pairing renormalization group introduced in \cite{metlitski15} was applied to the bilayer system.

To further this connection, we supplement (\ref{flow1}) with the equation,
\begin{eqnarray}
\frac{d\lambda_-}{dl} = \gamma \lambda_-,\label{flow2}
\end{eqnarray}
which follows directly from the definition of $\lambda_-$.  Equations (\ref{flow1}) and (\ref{flow2}) then form a system of coupled first-order differential equations for which the solutions can be visualized by flow diagrams in the $V, \lambda_-$ plane.  These diagrams, two of which are shown in Fig.~\ref{flow}, allow us to analyze the solutions of this equation in ways that are not immediately apparent starting with the gap equation (\ref{gap_equation}).  For that reason, in what follows we will focus first on the local approximation.  Later we will compare the exact numerical result with the local approximation results in order to verify that, while there are some quantitative differences, the local approximation appears to capture the qualitative behavior of this problem well.

\begin{figure}[t]
  \begin{center}
		 \includegraphics[width=3.3in]{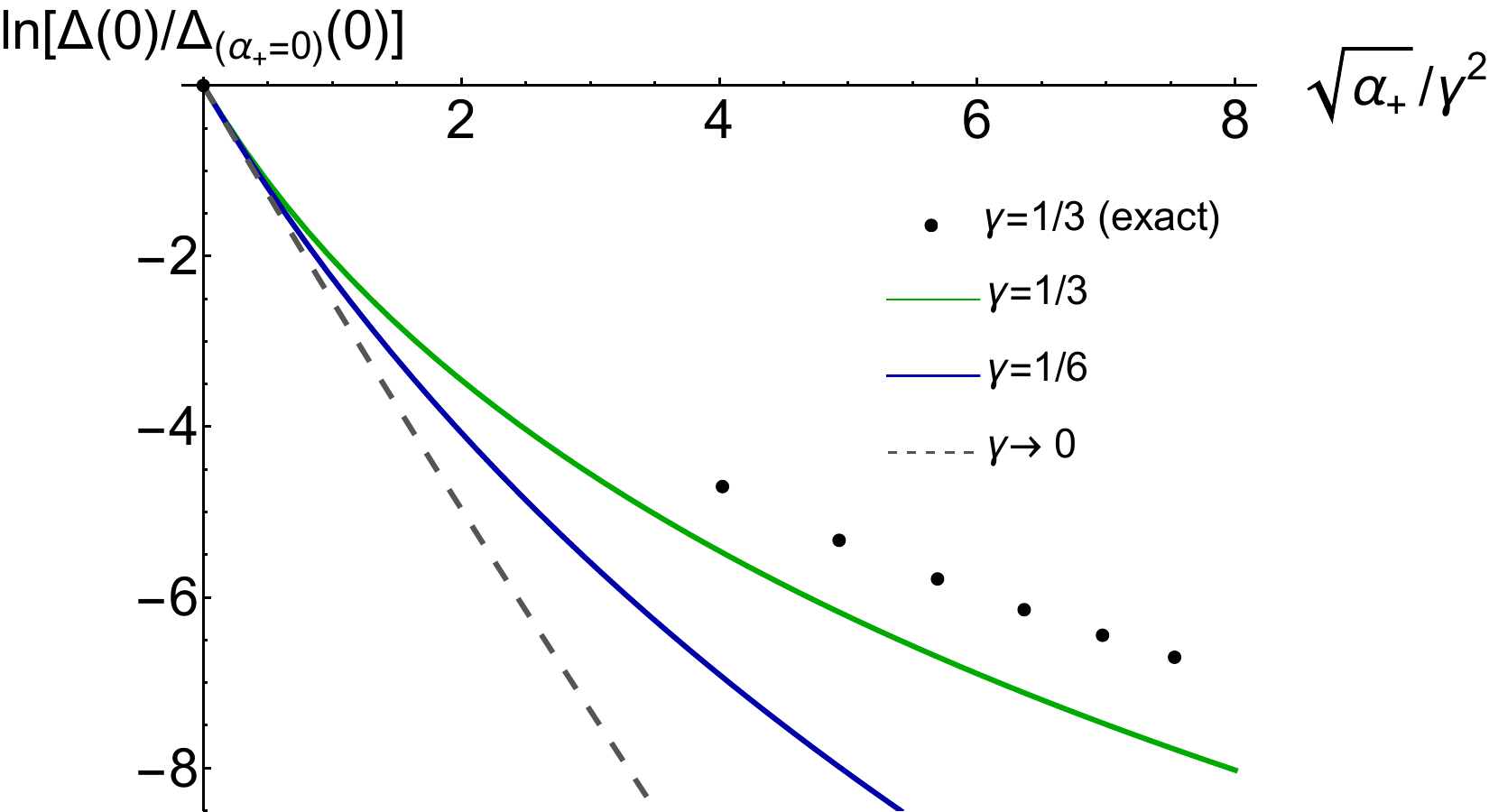}
	  \caption{(Color online). Log of the ratio of zero frequency gap with pair breaking to that without pair breaking as a function of the pair breaking parameter $\sqrt{\alpha_+}/\gamma^2$ for different values of $\gamma$.  Local approximation results are shown for $\gamma = 1/3$ (green line) and 1/6 (blue line), as well as the limit $\gamma \rightarrow 0$ (dashed line).  Exact numerical results for $\gamma = 1/3$ are also shown.} \label{sf}
   \end{center}
\end{figure}

Before proceeding we note some features of this flow diagram which will play important roles in what follows.  The UV boundary condition 
\begin{eqnarray}
V(\omega_0) = -\alpha_- + V_0;\ \ \ \lambda_-(\omega_0) = \alpha_-,
\end{eqnarray}
gives us a starting point in the flow diagram corresponding to the point where the flow parameter is $l = \ln \frac{\omega_0}{\omega} = 0$.  We see that in the limit of large layer spacing, for which $\alpha_- \ll 1$, provided $V_0 > -\sqrt{\alpha_+}$, the solution always flows to the trajectory which starts at the point $(V,\lambda_-) = (+\sqrt{\alpha_+},0)$.  This gives a universal form for $V(\omega)$ in this limit, the details of which are discussed in Sec.~\ref{lls}.  

We see that if $\alpha_-$ is kept small, but $V_0$ is varied in such a way that it becomes more and more negative, eventually, once $V_0 < -\sqrt{\alpha_+}$, there will be an abrupt crossover to a pairing phase with a much bigger energy gap.  We view this crossover as being between a ``gauge pairing" phase, in which the pairing interaction is dominated by long-wavelength low-frequency gauge fluctuations, to a paired state that is much closer to a conventional BCS superconductor with the pairing due primarily to a short-range nonsingular attractive $V_0$.  It is natural to take the flow trajectory which starts at $(V,\lambda_-) = (-\sqrt{\alpha_+},0)$ as the crossover point, for which $V(\omega)$ flows initially to $0$, just touching this value, before flowing to $-\infty$ indicating the formation of a paired state.  This crossover is further discussed in Sec.~\ref{xover}.

Aside from providing a link to the renormalization group approach, one appealing feature of the local approximation is that it allows us to obtain essentially analytic results for $\Delta(\omega)$ and its dependence on $\alpha_-$, $\alpha_+$, and $V_0$.  In addition to being of value in their own right, these approximate $\Delta(\omega)$ solutions can be used as starting points to iterate the exact gap equation, which we find then rapidly converges.  By performing such numerical iteration we are able to compare the analytic local approximation results with the results of actually solving the gap equation.

\section{Large Layer Spacing Limit} \label{lls}

In this section, we examine the limit of large layer spacing more closely.  As noted above, in this limit the coupling constant $\alpha_- << 1$, and the flow in Fig.~\ref{flow}  begins asymptotically close to the $\lambda_- = 0$ line.  In the limit of large layer spacing, we expect the coupling $V_0$ to be very small in magnitude, but we see here that if $V_0 < -\sqrt{\alpha_+}$, the flow changes drastically, reflecting the crossover from gauge pairing to more BCS-like pairing (discussed in the next section).

\begin{figure}[t]
  \begin{center}
	\centerline{Local Approximation\ \ \ \ \ \ \ \ \ \ \ \ \ \ \ \ \ \ \ \ \ \ \ \ \ \ Exact\ \ \ \ \ \ }
	 \includegraphics[width=1.7in]{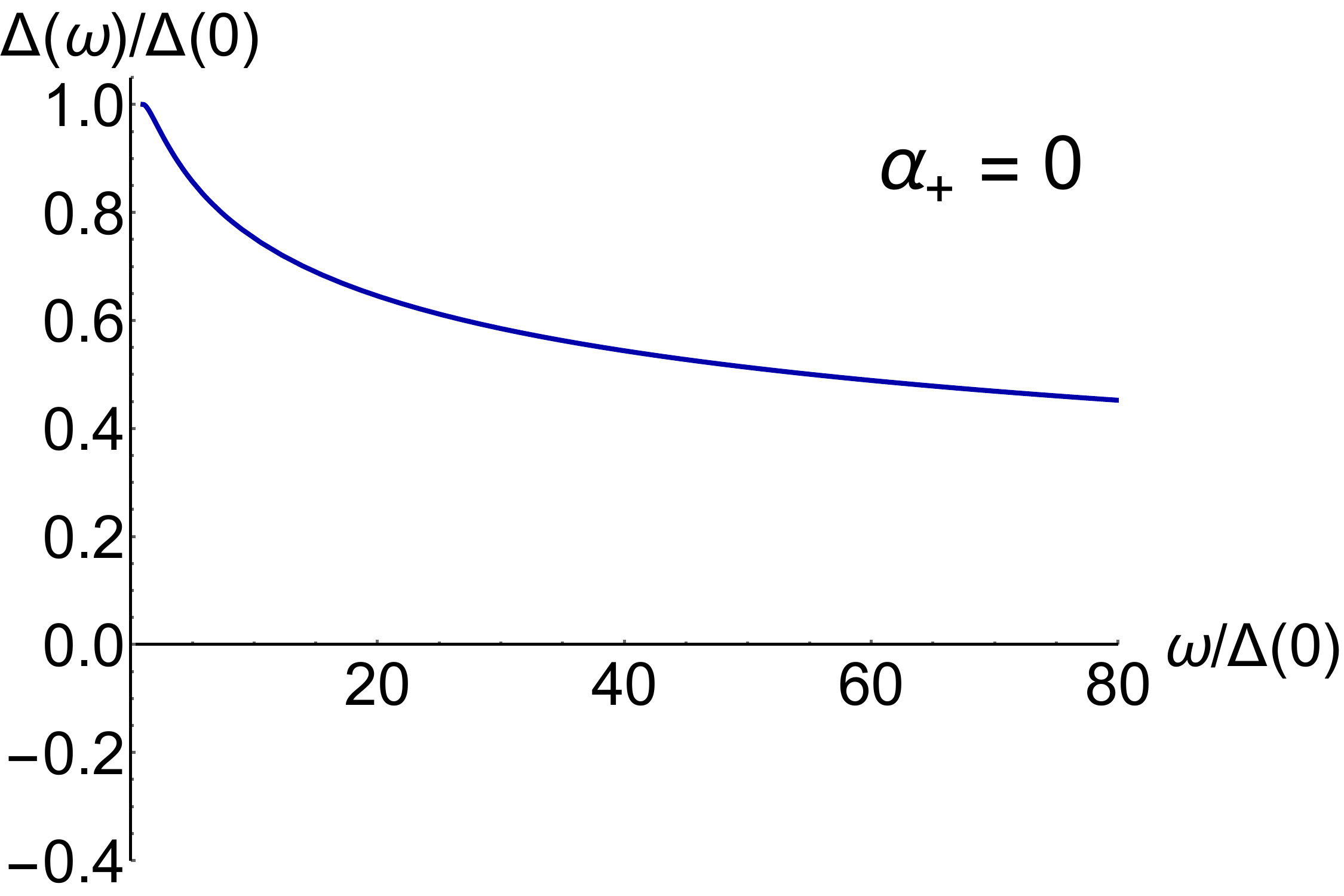}~~
	 \includegraphics[width=1.7in]{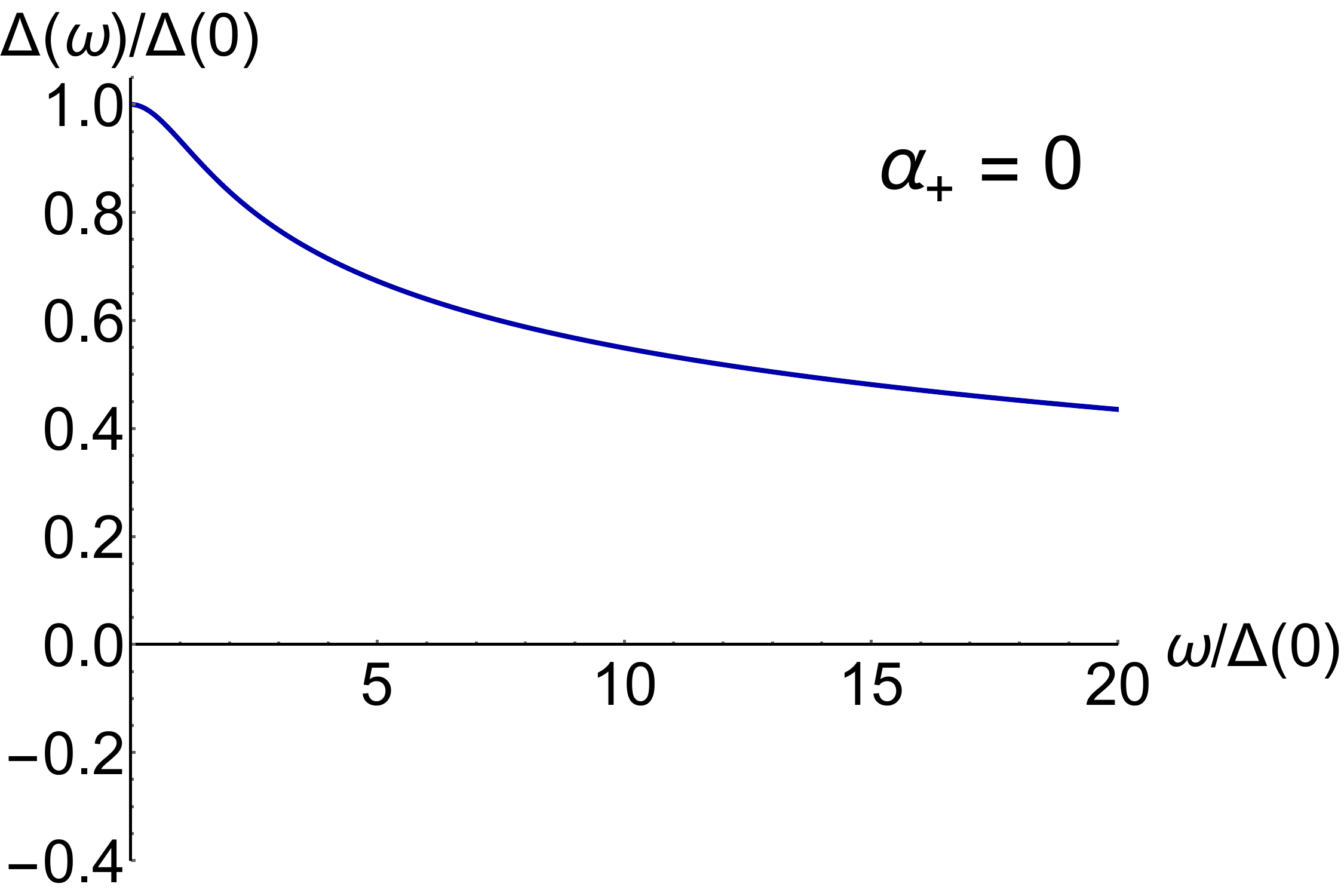}\\ \vskip .2in
		\includegraphics[width=1.7in]{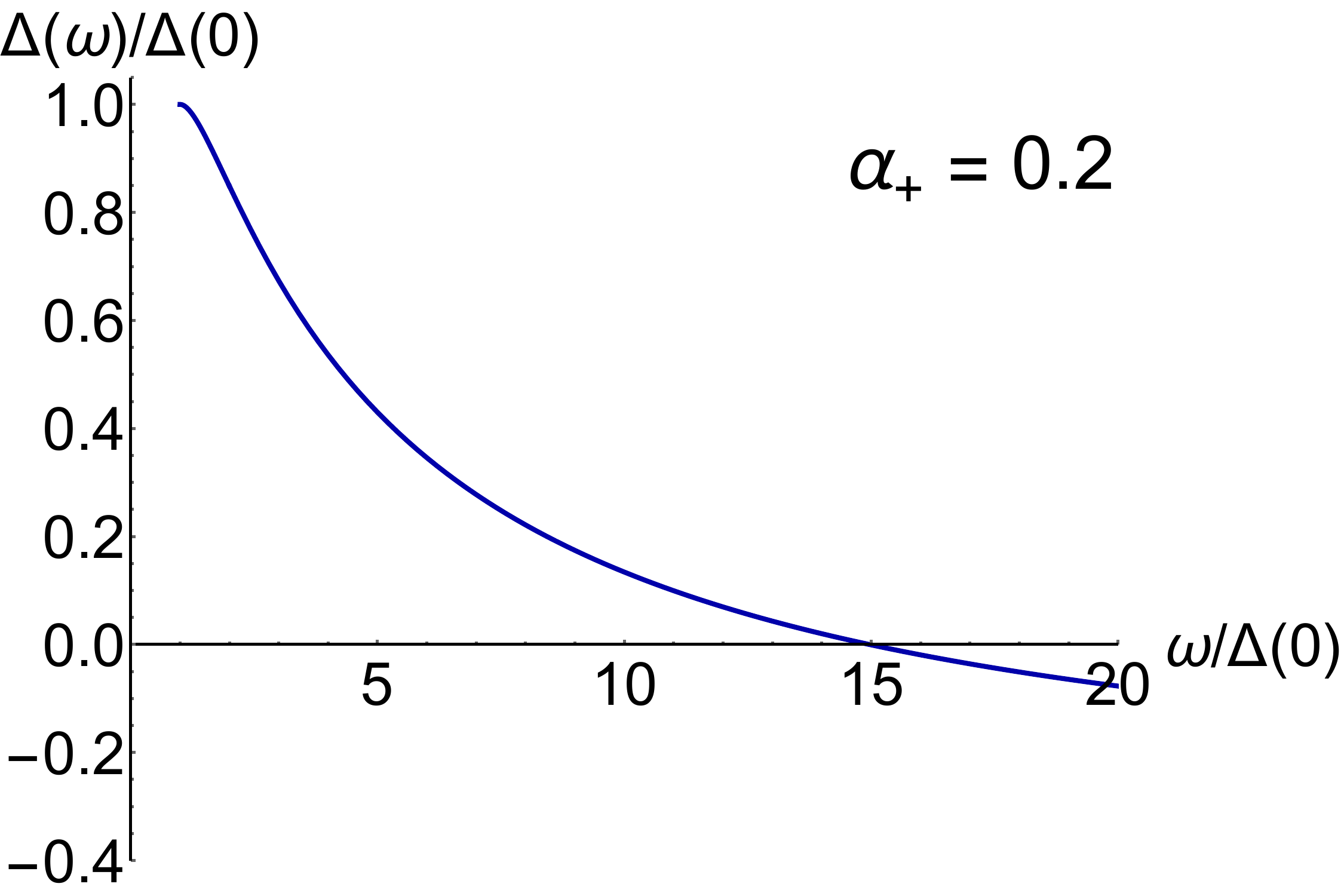}~~
	 \includegraphics[width=1.7in]{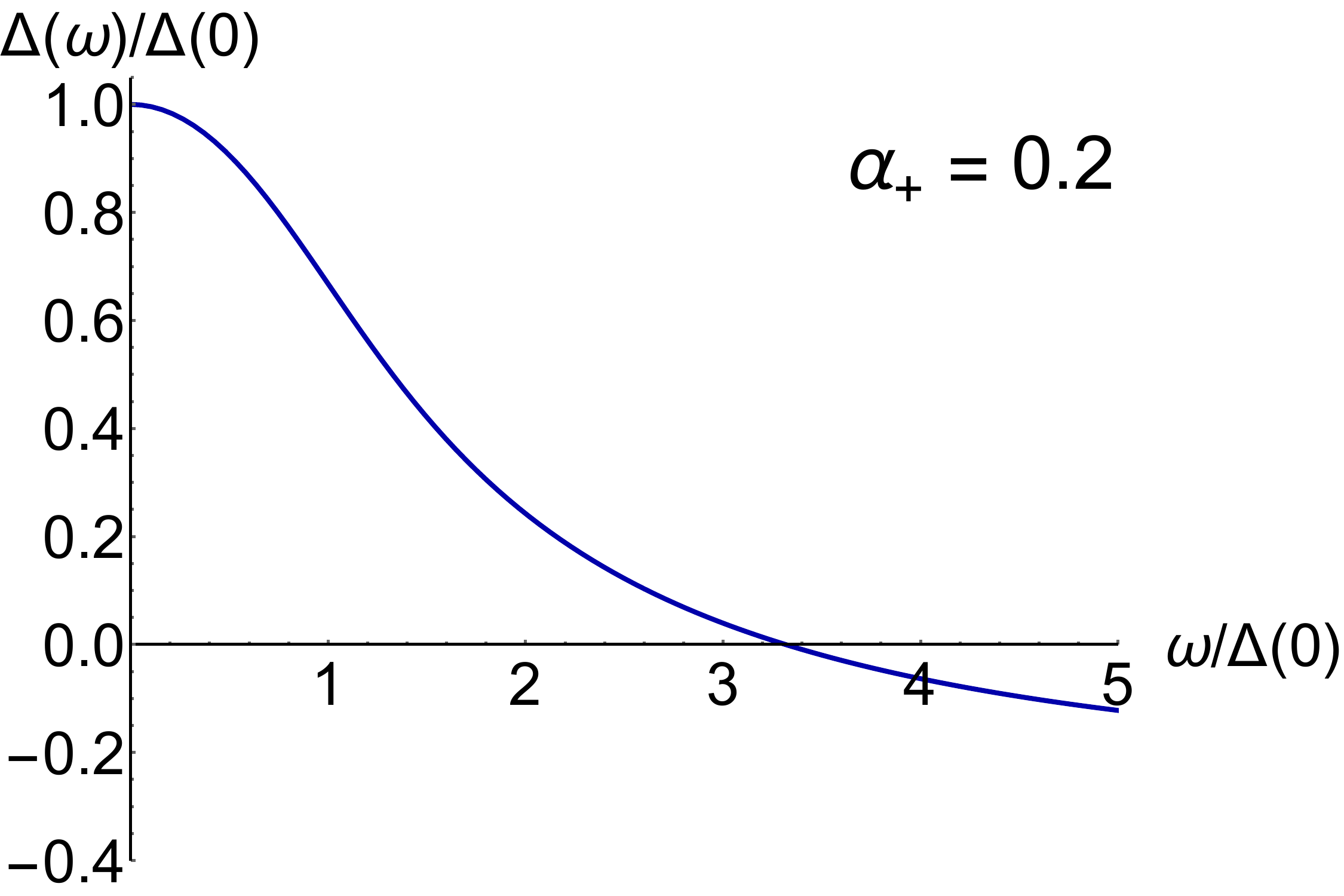}\\ \vskip .2in
		\includegraphics[width=1.7in]{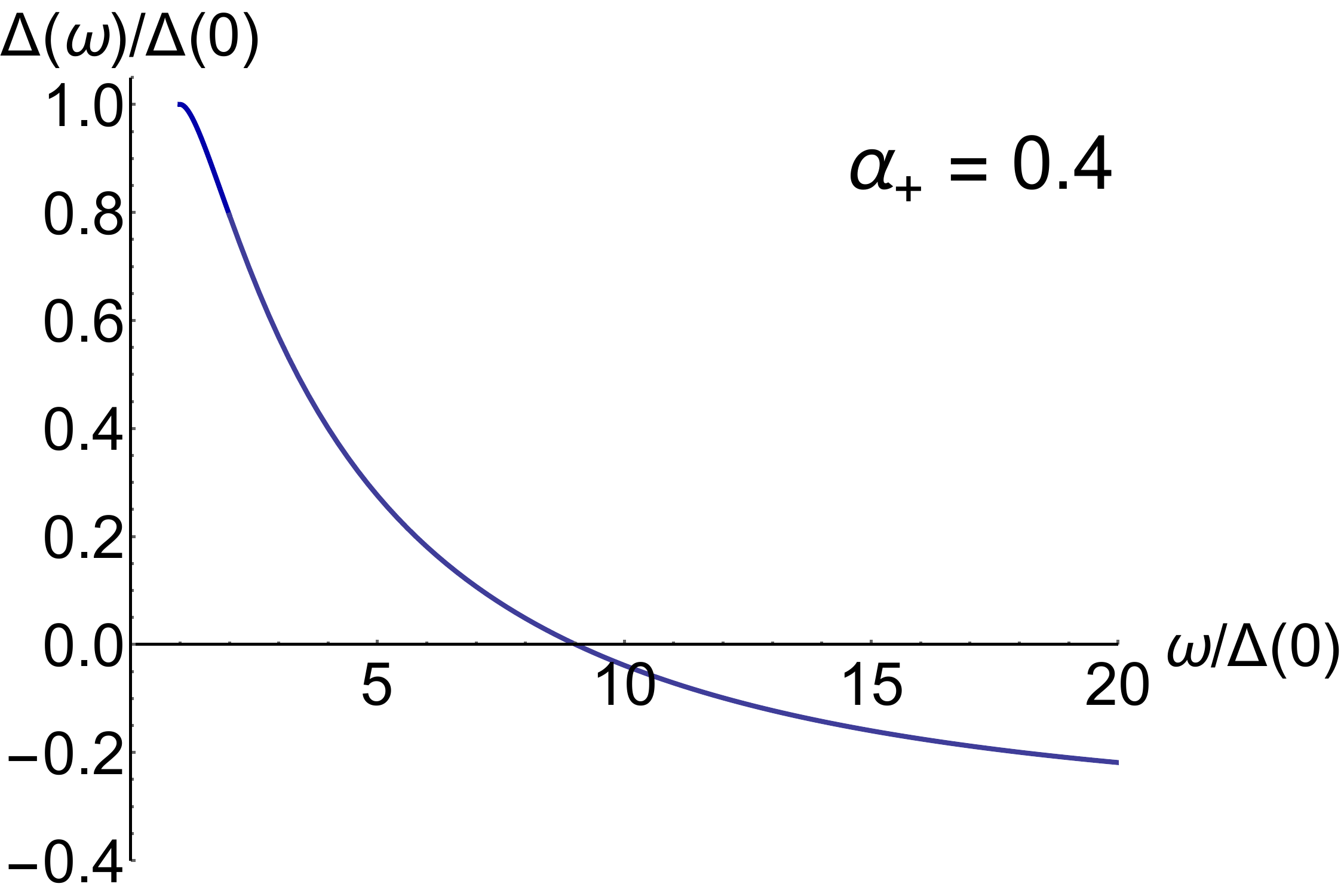}~~
	 \includegraphics[width=1.7in]{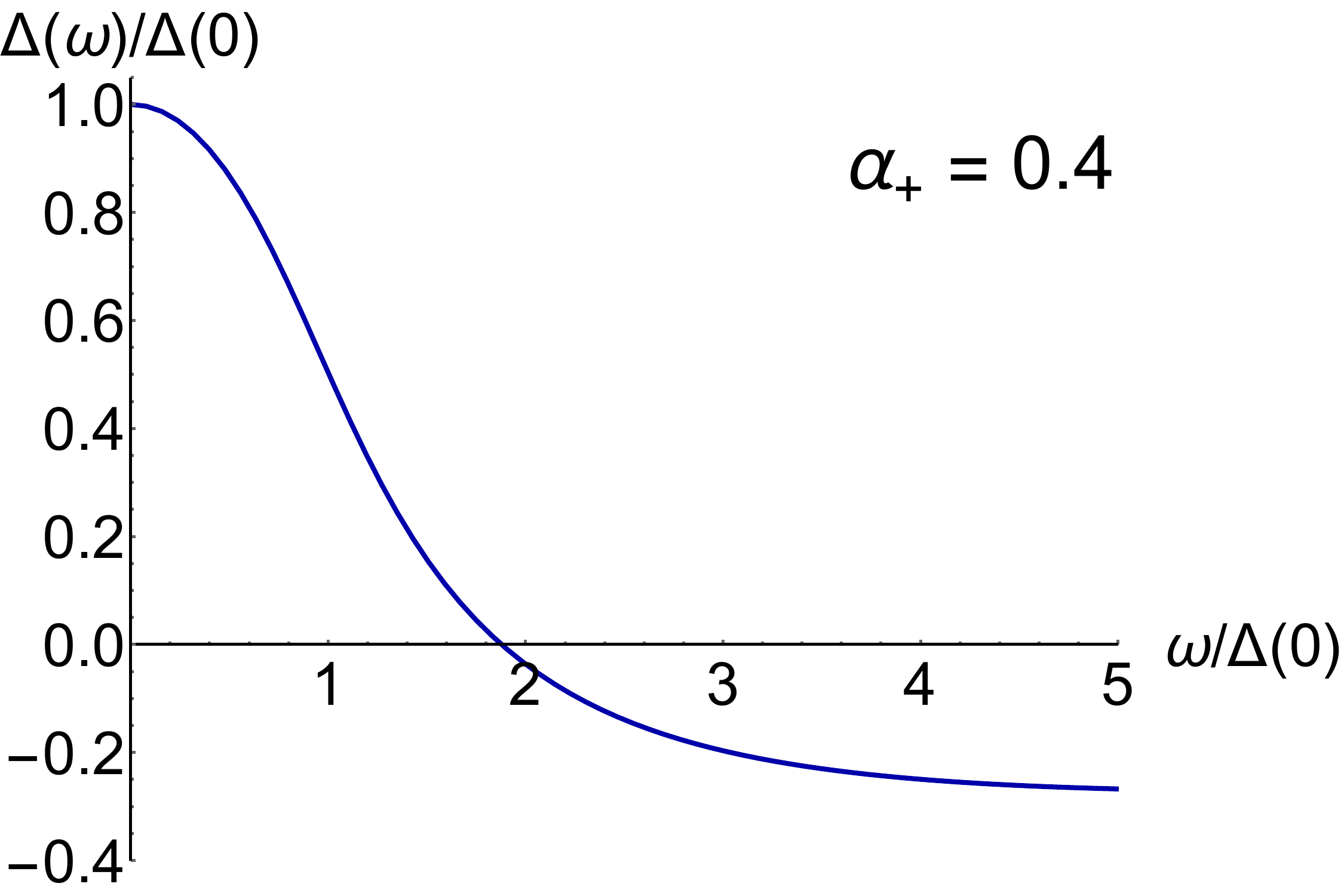}
  \caption{(Color online). $\Delta(\omega)$ vs. $\omega$ for $\alpha_+ = 0, 0.2, 0.4$.  Results are shown for the local approximation, as well as the exact solution of the gap equation obtained by iterating the gap equation, starting with the local approximation solution.  Both the local approximation and exact results show the sign change in $\Delta(\omega)$ when $\alpha_+ > 0$, characteristic of the combination of pairing and pair breaking with different frequency dependencies as discussed in the text.} \label{smallam}
   \end{center}
\end{figure}

This $V_0 = -\sqrt{\alpha_+}$ unstable fixed point for $\alpha_-=0$ corresponds to the phase transition between the unpaired and paired state as a function of $V_0$ when $\alpha_- = 0$.  In this case, we are simply analyzing the problem of BCS pairing due to an attractive nonsingular $V_0$ interaction, in the presence of marginally singular pair breaking due to $\lambda_+$.     This singular pair breaking leads to a qualitative change in the usual BCS pairing problem.  A well-known feature of BCS pairing is that it occurs for {\it any} attractive interaction strength.  But here, because of the singular pair breaking due to the in-phase gauge fields, there is a critical interaction strength needed, as shown earlier in \cite{metlitski15,bonesteel99}, with the flow equations appearing here closely related to those in \cite{metlitski15}.

As $\alpha_-$ becomes finite, this phase transition becomes a crossover.  Again, if the layer spacing is large we expect $V_0$ to be negligibly small, but as the layers are brought together the variation of $V_0$ can lead to a crossover, which may correspond to the experimental observation of a gap turning on continuously as $d/l_0$ is decreased below a critical value ($\sim 1.6$ in experiment).   First we consider the $\alpha_- << 0$ limit, and assume that $V_0 > -\sqrt{\alpha_+}$.  In this case, the flow diagram makes clear that the solution flows to the trajectory that starts at the point $\lambda_-=0$, $V_0 = +\sqrt{\alpha_+}$.  We therefore continue to see scaling behavior even with $\alpha_+$ present.  For the universal scaling curve, one can find $V(\omega)$ and use it to analytically obtain the zero frequency gap $\Delta(0)$, as well as $\Delta(\omega)$ using (\ref{vtod}).  The procedure for finding $V(\omega)$ for this case is described in the Appendix.

To address the question of how effective the in-phase gauge fluctuations are for pair breaking, we show in Fig.~\ref{sf} the ratio of the zero frequency gap with $\alpha_+$ present to that in the absence of pair breaking for different values of $\gamma$ within the local approximation, as well as numerical results for the $\gamma = 1/3$ case. It is apparent from this figure and is shown explicitly in the Appendix, that in the limit of small $\gamma$, the zero frequency gap has the form
\begin{eqnarray}
\Delta(0) = e^{(-2.566\cdots)\frac{\sqrt{\alpha_+}}{\gamma^2}}\Delta_{\alpha_+ = 0}(0)
\end{eqnarray} \label{ratio}
where 
\begin{eqnarray}
\Delta_{(\alpha_+ = 0)}(0) = \left(\frac{{0.6917 \cdots}}{\gamma}\right)^{1/\gamma} \alpha_-^{1/\gamma} \omega_0
\end{eqnarray}
is the zero frequency gap in the absence of pair breaking. 

From these expressions, we see that in the absence of pair breaking, for which $\alpha_+ = 0$, for $\gamma =1/3$ and $\alpha_ \sim (l_0/d)^{2/3}$ we have $\Delta_{(\alpha+ = 0)}(0) \sim \omega_0 (l_0/d)^2$, which is the well-known result that $\Delta(0)$ is proportional to inverse layer spacing squared for this case \cite{bonesteel96}.   When $\alpha_+$ is now turned on, we can identify a pair-breaking parameter $\sqrt{\alpha_+}/\gamma^2$, which serves as a dimensionless measure of the effect of the pair breaking due to these out-of-phase gauge fluctuations.   This result makes physical sense and reflects the fact that for smaller $\gamma$, and hence less singular pairing fluctuations, the pair breaking is more effective.   It further shows for the physically relevant case of $\alpha_+ \sim 1$ and $\gamma = 1/3$ pair breaking is quite strong.  The observation that this pair breaking is so strong, despite the fact that the pair-breaking fluctuations are less singular than the pairing fluctuations, may account for the fact that the predicted inverse layer spacing squared dependence of the energy gap is not seen experimentally.  While, as a matter of principle, for a perfectly clean sample at low enough temperature, a paired state may always form even at large layer spacing, in practice disorder will suppress the transition once the layer spacing is large enough that the pairing gap becomes comparable to the disorder width.

Figure \ref{smallam} shows the frequency-dependent gap $\Delta(\omega)$ for various values of $\alpha_+$ in the small $\alpha_-$ limit.  Results are shown both for the local approximation and the exact numerical solution, obtained by using the local approximation as a starting solution and iterating the gap equation.  The results show that the effect of the singular pair-breaking term on the frequency dependence of the energy gap is significant.  It leads to a change in sign in $\Delta(\omega)$ which, in the local approximation, occurs when the $V$ trajectory on the flow diagram passes through 0.  Such a sign change is similar to that which occurs in other pairing systems with a combination of attractive and repulsive contributions to the effective coupling, each having different frequency dependencies --- in our case ``fast" pair-breaking in-phase fluctuations and ``slow" pairing out-of-phase fluctuations.  The possible physical consequences of this sign change have recently been discussed in \cite{christensen21} (see also \cite{pimenov22}) in the context of taking Coulomb repulsion into account in conventional BCS theory \cite{morel62} (in that case the Coulomb repulsion is ``fast" and the pairing due to phonons ``slow").  This zero in $\Delta(\omega)$, corresponds to a dynamical vortex in its analytic structure on the imaginary axis of the type discussed in \cite{christensen21}.  As in that case, upon analytic continuation, this implies the real and imaginary parts of the gap must (in general separately) change sign for some finite frequencies on the real axis.  While this vanishing of the real and imaginary parts of the gap may be experimentally observable for electronic superconductors as discussed in \cite{christensen21}, it is less clear how this could be observed in the present case given that the particles undergoing pairing are composite fermions which are highly nonlocal objects when viewed in terms of physical electrons.   Nonetheless, we believe the appearance of a dynamical vortex in this pairing problem can be viewed as a ``singular" version of that discussed in \cite{christensen21,pimenov22}.

\section{Crossover from gauge pairing to BCS pairing}\label{xover}

The significant effect of in-phase pair breaking discussed above suggests the following scenario for the formation of an observable bilayer quantum Hall state using the model interaction (\ref{model}).  The nonsingular parameter $V_0$, which unlike $\alpha_+$ and $\alpha_-$ depends on the angular momentum channel $l$, can be viewed as a semi-phenomenological parameter that describes the short-range part of the interlayer interaction and can be used to drive a crossover from the gauge pairing ($V_0$ irrelevant) regime to BCS ($V_0$ driven) pairing regime.  We identify the crossover from these two different families of flow trajectories as the trajectory that starts at the point $V_0 = -\sqrt{\alpha_+}$, $\lambda_- = 0$ and just touches the $V=0$ axis without crossing it (bottom red flow in both diagrams shown in Fig.~\ref{flow}). 

\begin{figure}[t]
  \begin{center}
	 \includegraphics[width=3in]{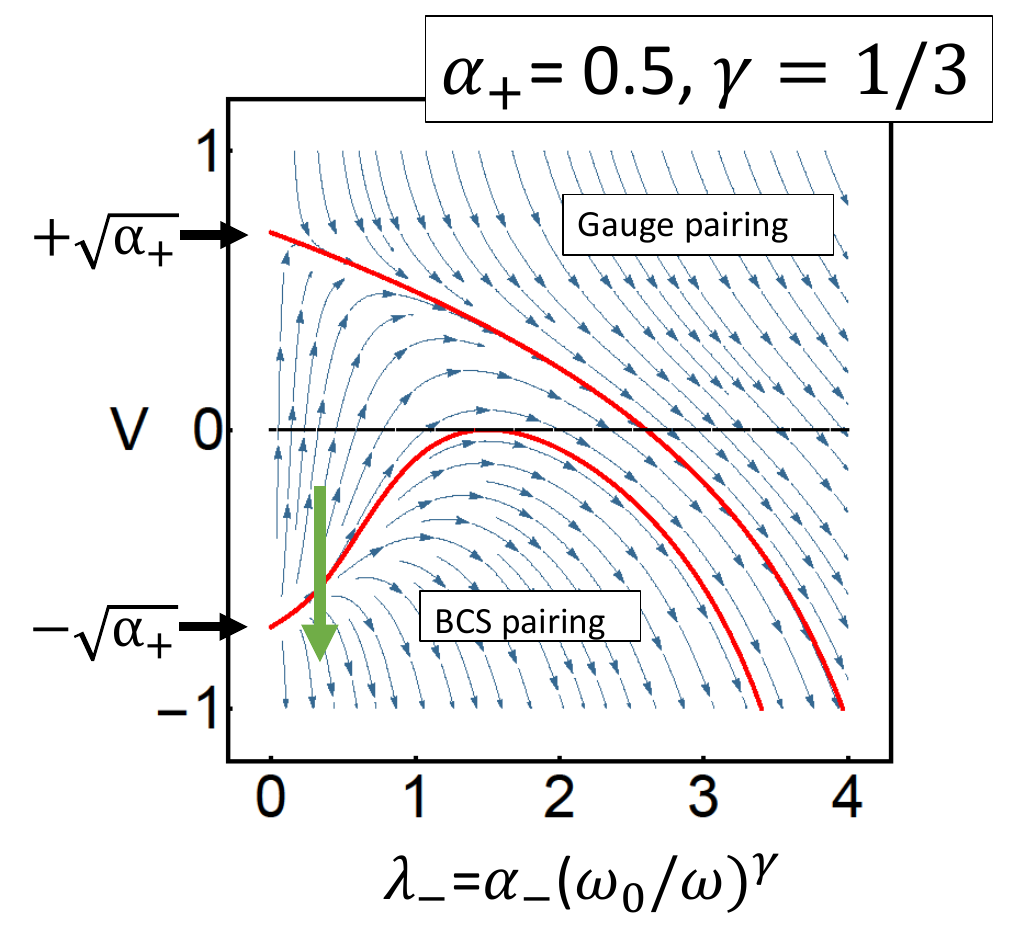}
  \caption{(Color online)  $V, \lambda_-$ flow diagrams for the case $\alpha_+ = 0.5$ and $\gamma =1/3$.  The green arrow shows initial states for the case keeping $\alpha_- = 0.3$ and varying $V_0$ so that it passes through the crossover from the gauge pairing regime, in which pairing is dominated by long-wavelength low energy gauge fluctuations, and strongly suppressed by pair-breaking fluctuations, to BCS pairing in which pairing is dominated by an attractive nonsingular pairing interaction.} \label{flow_co}
   \end{center}
\end{figure}

\begin{figure}[b]
  \begin{center}
	\centerline{Local Approximation\ \ \ \ \ \ \ \ \ \ \ \ \ \ \ \ \ \ \ \ \ \ \ \ \ \ Exact\ \ \ \ \ \ }
	 \includegraphics[width=1.7in]{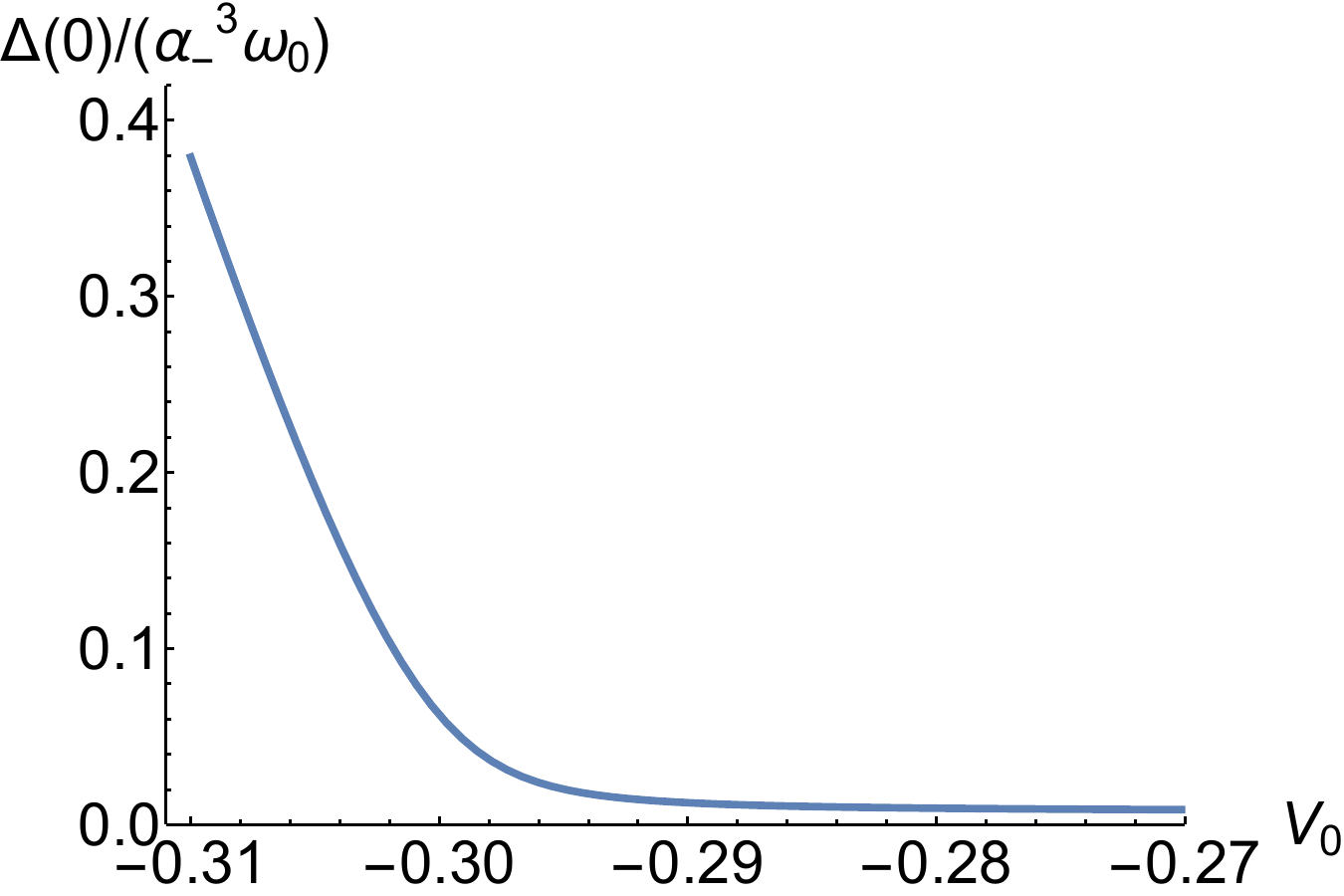}~~
	 \includegraphics[width=1.7in]{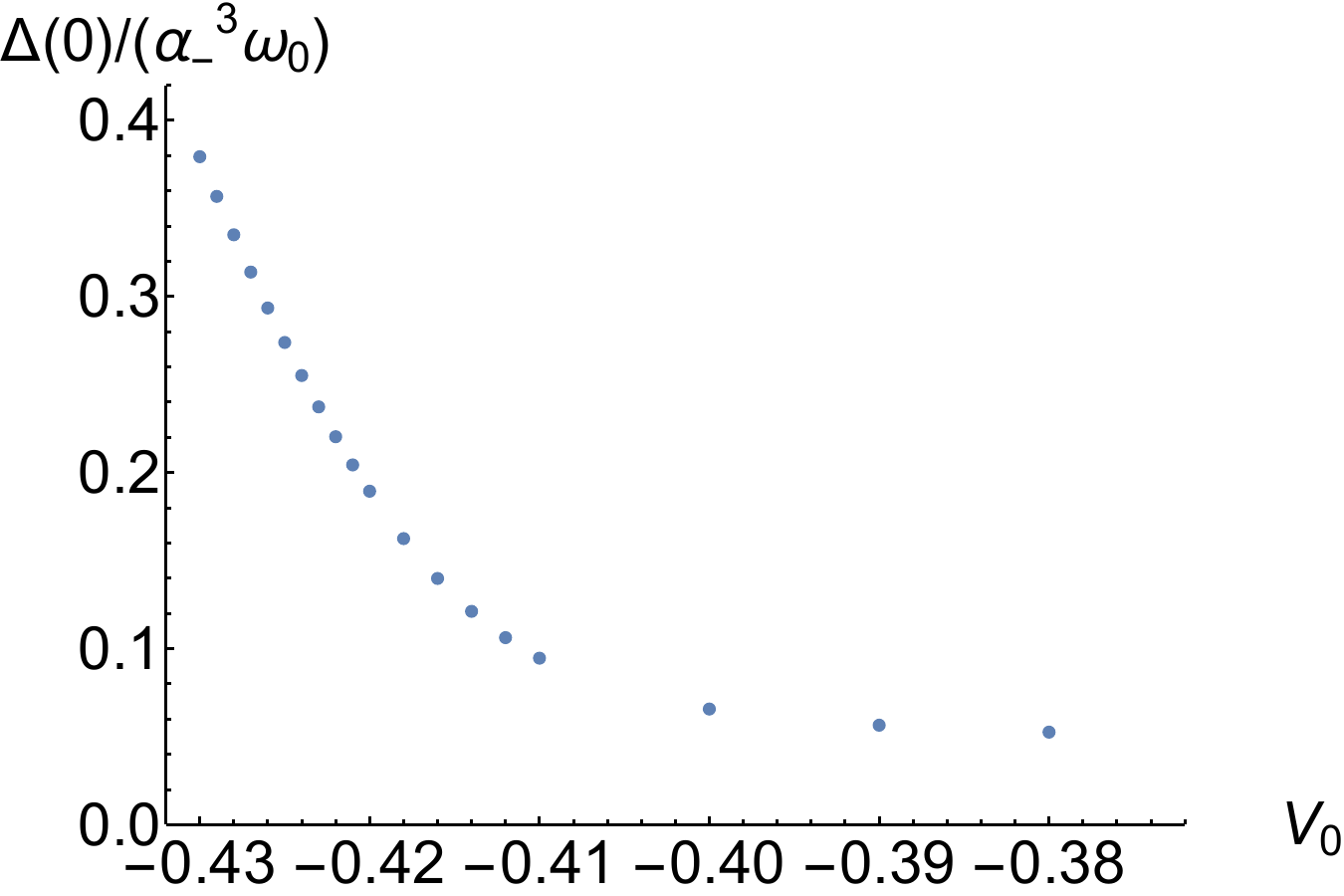}
  \caption{(Color online).  $\Delta(0)$ vs. $V_0$ for $\alpha_+ = 0.5$ and $\alpha_- = 0.3$ for the local approximation result obtained analytically and for the exact solution obtained numerically.  The local approximation results correspond to the trajectory shown in green in Fig.~\ref{flow_co}}\label{gapvsv}
   \end{center}
\end{figure}

As the layers are brought together the coupling $\alpha_-$ becomes larger, and we can imagine that the parameter $V_0$ becomes negative enough, i.e. provides strong enough nonsingular pairing, to drive this crossover.  We further assume that the leading instability occurs within the appropriate $p$-wave channel, as suggested by more detailed RPA calculations \cite{isobe17,lotrivc23}. It is natural to suppose that, within the framework of the model discussed here, this is precisely what happens when the transition from bilayer composite fermion metal to paired state is observed in experiment.

\begin{figure}[t]
  \begin{center}
	\centerline{Local Approximation\ \ \ \ \ \ \ \ \ \ \ \ \ \ \ \ \ \ \ \ \ \ \ \ \ \ Exact\ \ \ \ \ \ }
	 \includegraphics[width=1.7in]{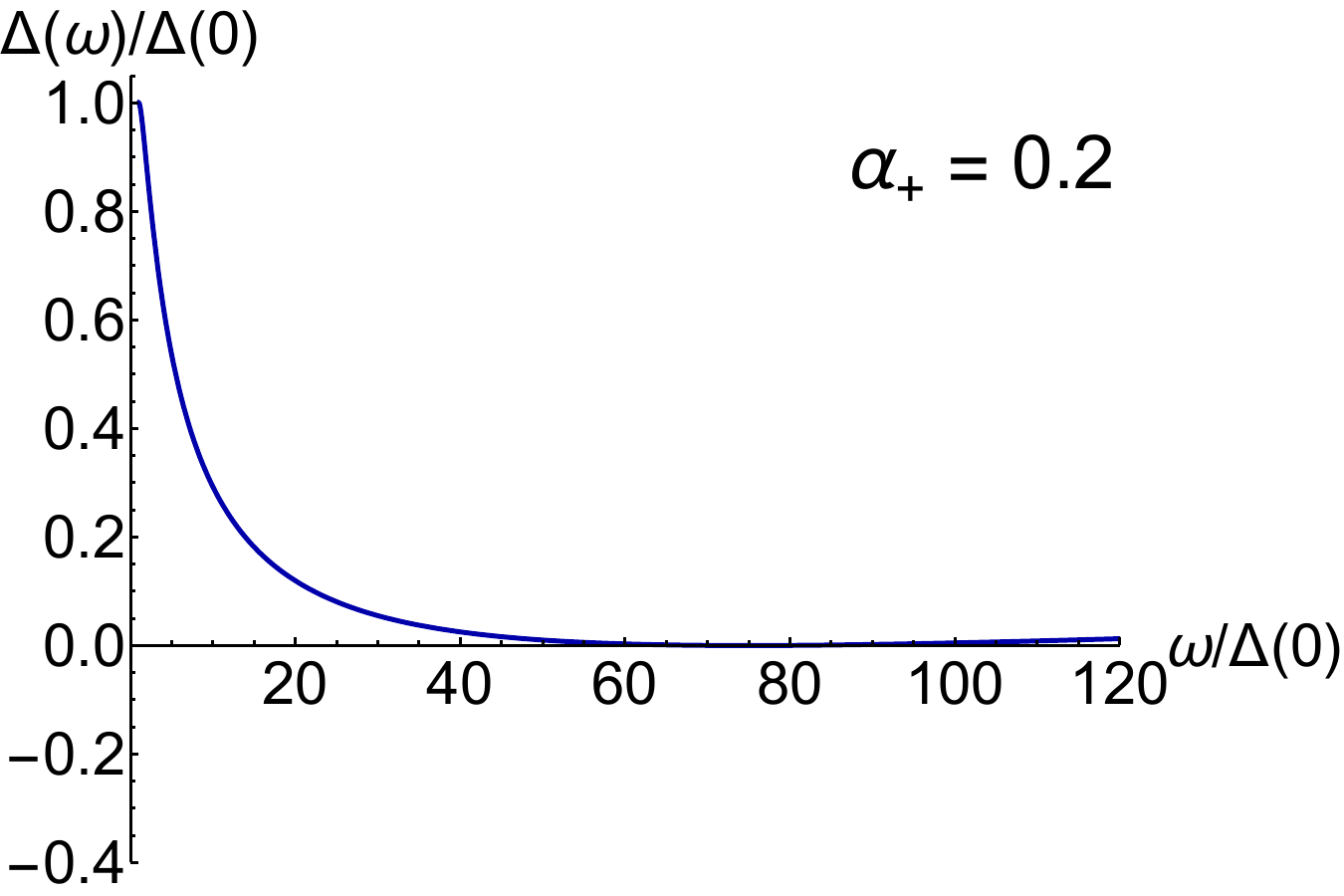}~~
	 \includegraphics[width=1.7in]{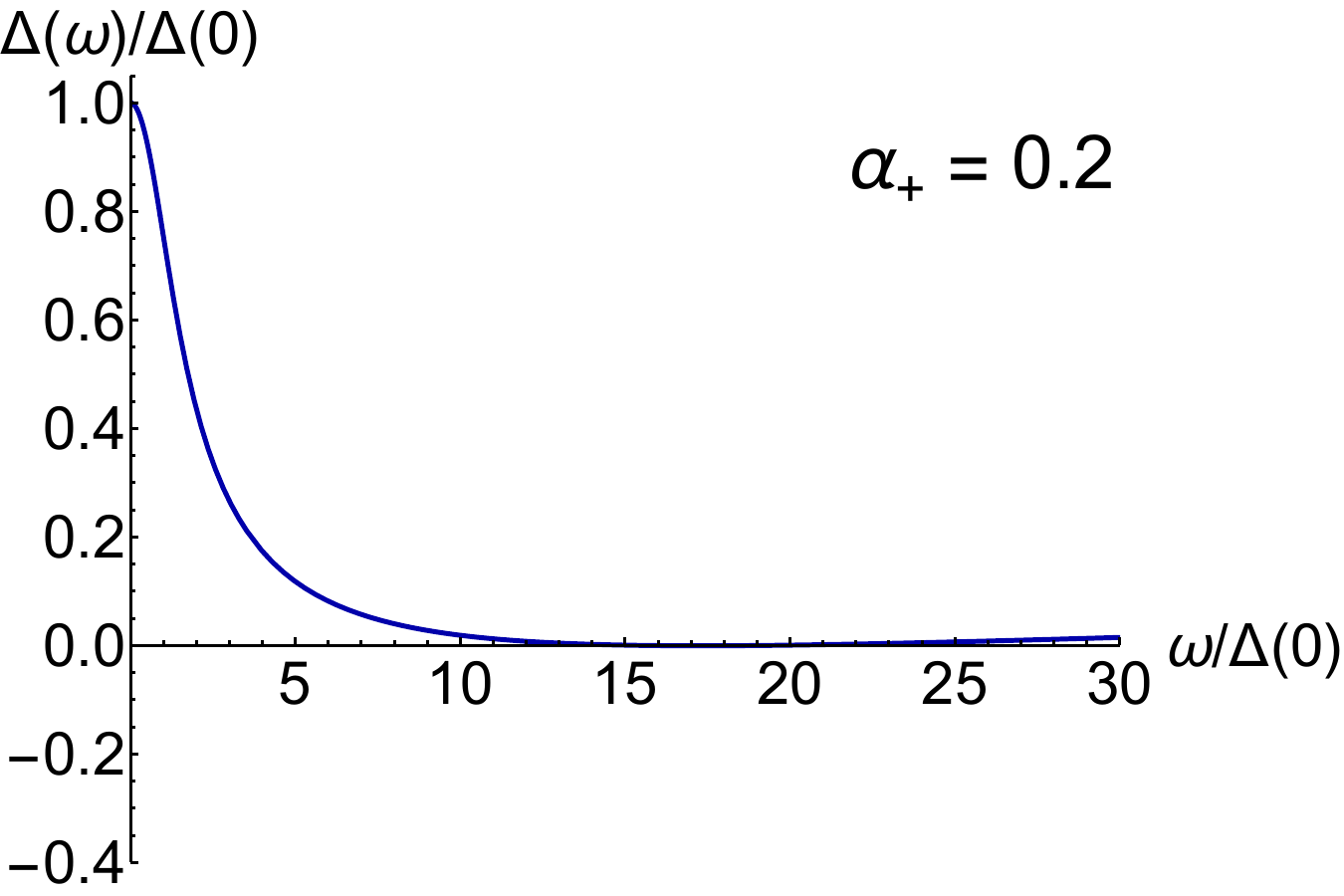}\\ \vskip .2in
		\includegraphics[width=1.7in]{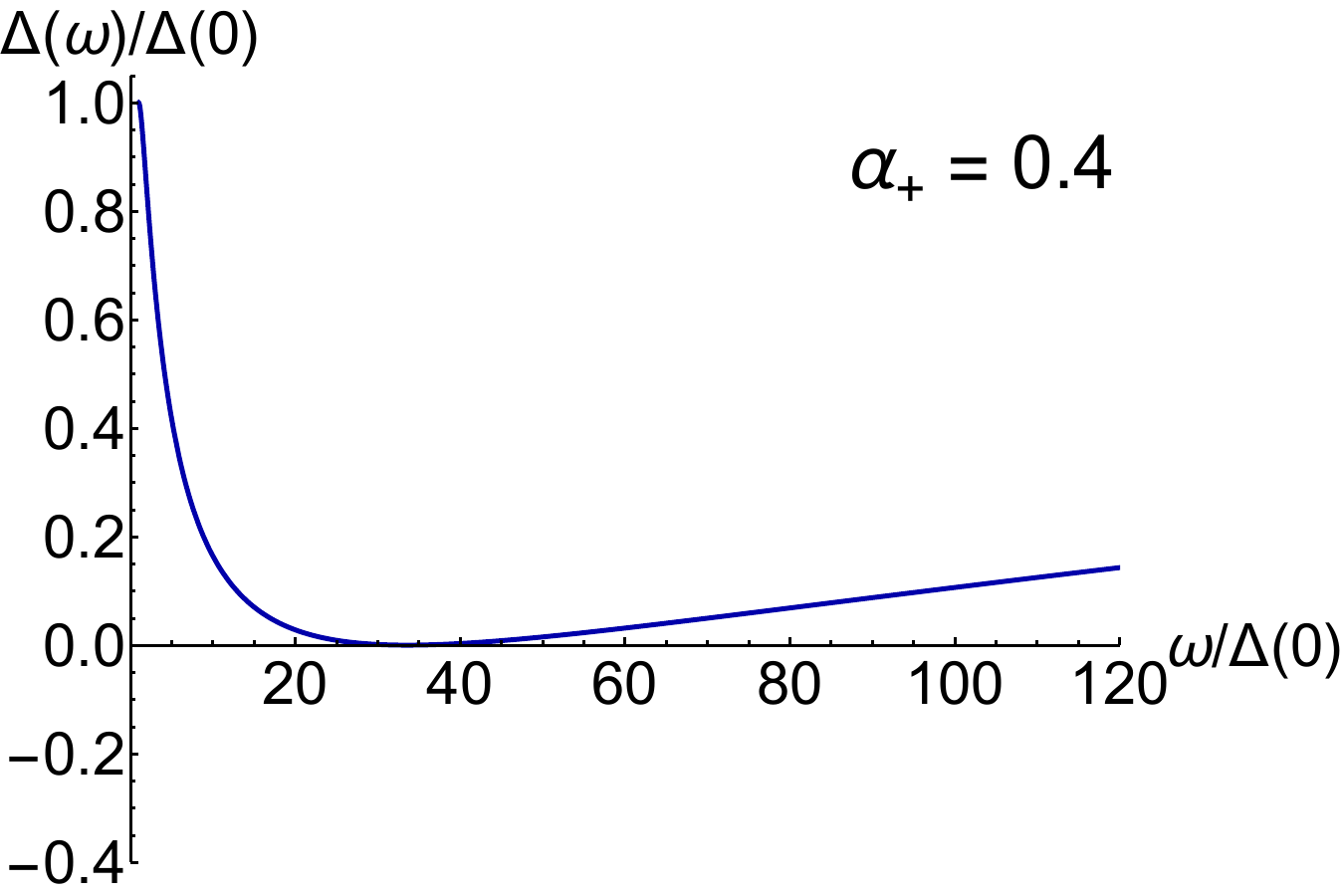}~~
	 \includegraphics[width=1.7in]{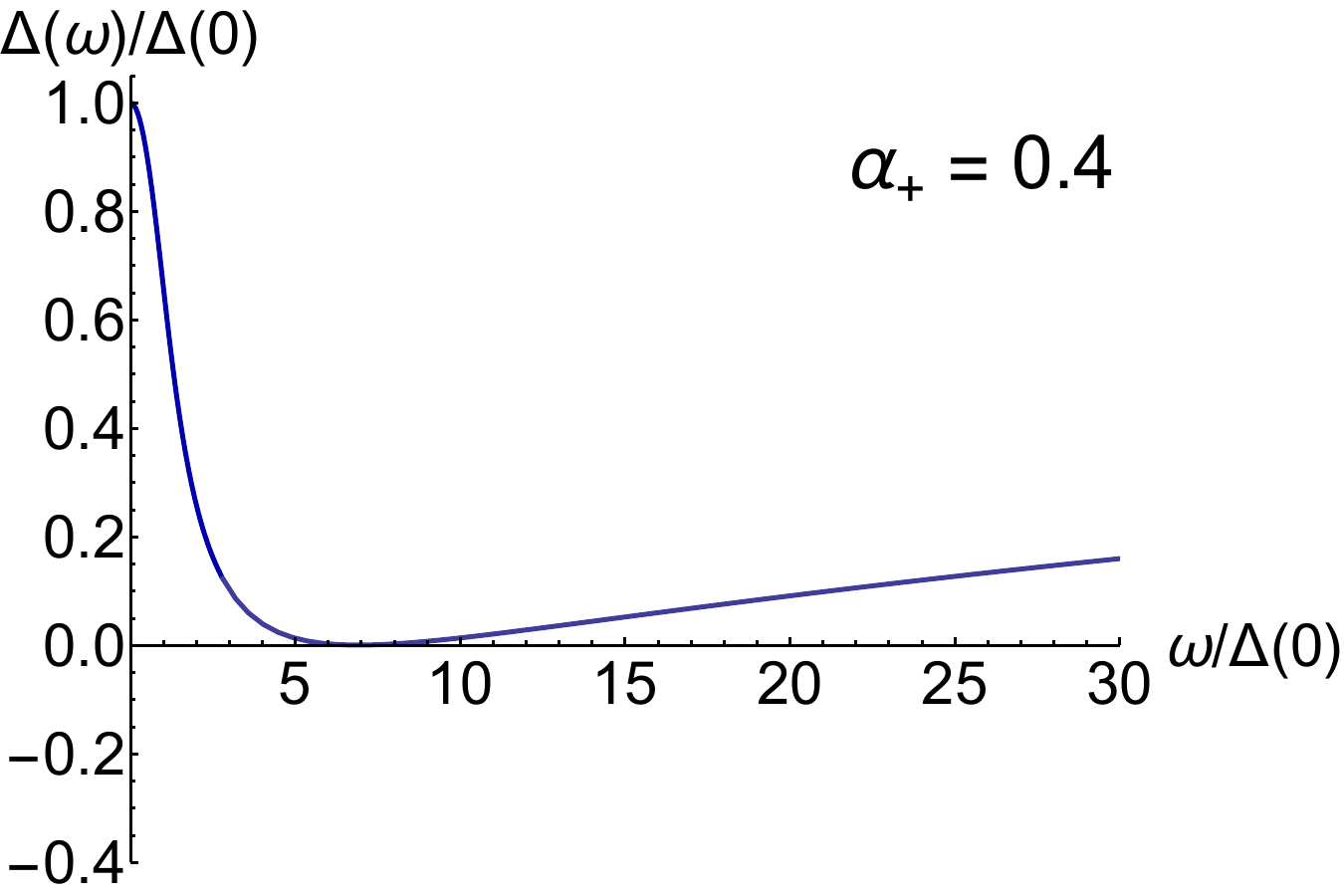}
  \caption{(Color online).  $\Delta(\omega)$ vs. $\omega$ for $\alpha_+ = 0.2, 0.4$ at the crossover from gauge paring to BCS pairing.  Results are shown for the local approximation, as well as the exact solution of the gap equation obtained by iterating the gap equation, starting with the local approximation solution.  All results are for $\alpha_- =.1$.  For the local approximation results  $V_0 = -.252657$ for $\alpha_+ = .2$ and $V_0 = -.493164$ for $\alpha_+ = .4$.  For the exact results $V_0 = -.28$ for $\alpha_+ =.2$ and $V_0 = -.6316$ for $\alpha_+=.4$} \label{co}
   \end{center}
\end{figure}

An example of such a crossover is sketched in the $\alpha_+ = 0.5$ flow diagram shown in Fig.~\ref{flow_co}, with the initial flow conditions $\alpha_- = 0.3$ and $V_0$ being tuned through the crossover from gauge pairing to BCS pairing (green arrow).  The dependence of the corresponding energy gap $\Delta(0)$ on $V_0$ is shown in Fig.~\ref{gapvsv}, both for the local approximation and for the numerically obtained exact solutions of the gap equation.  Both solutions show a steep increase in $\Delta(0)$ as the crossover occurs, though the value of $V_0$ for the crossover is, of course, different for the approximate and exact solutions.

We can take the crossover value for $V_0$ to be that value for which the trajectory in the flow diagram just touches the $V = 0$ axis.  It is for this value that the sign change in the gap function just disappears, signaling a crossover from strongly suppressed gauge pairing to more BCS-like pairing, which, as $V_0$ becomes more negative, will lead to a substantial gap persisting all the way up to the cutoff $\omega_0$.   

Within the local approximation, this crossover value of $V_0$ corresponds to the trajectory which starts at $\lambda_- = 0$ and $V_0 = -\sqrt{\alpha_+}$.  These solutions can be found analytically within the local approximation (see Appendix) and then used following (\ref{vtod}) to find the ``crossover" form of $\Delta(\omega)$, and are shown for the local approximation in Fig.~\ref{co}.  Note that these gap functions no longer change sign, but simply touch the $\Delta = 0$ axis at some finite imaginary frequency.  Results are also shown in Fig.~\ref{co} for the numerically obtained exact solution of the gap equation, where we again take our criteria for the crossover value of $V_0$ to be that which corresponds to $\Delta(\omega)$ just touches zero, without changing sign.  For the exact solution, it was necessary to adjust $V_0$ and $\alpha_-$, since the critical value of the transition is not determined analytically by the local approximation.  Nonetheless, we see qualitative agreement between the crossover gap functions for both the local approximation and the gap equation, suggesting that the local approximation captures the essential behavior of the solutions of (\ref{gap_equation}) for the model interaction (\ref{model}).

\section{Conclusions}
\label{conc}

In this paper, we have analyzed the solution to the $T=0$ gap equation for a model frequency-dependent interaction meant to describe the total $\nu=1/2+1/2$ bilayer quantum Hall state starting from the limit of large layer spacing.  The effective interaction contains a singular attractive interlayer pairing interaction, a less singular repulsive interlayer pairing interaction, and a non-singular coupling which we expect to depend on details of the system and also determine the pairing channel.   Using the local approximation we obtain essentially analytic approximate solutions of the gap equation, illustrating the very strong pair-breaking effect of the less singular in-phase gauge fluctuations.  We then argue that within this interlayer pairing scenario, the experimentally observed transition from two compressible bilayers to an incompressible bilayer quantum Hall state may reflect a crossover from a gauge pairing regime to a more conventional BCS-type pairing.

\acknowledgments
HD acknowledges useful discussions with Wai-Ga Ho.  The National High Magnetic Field Laboratory is supported by the National Science Foundation through NSF/DMR1644779 and by the State of Florida.

\onecolumngrid

\appendix*

\section{Flow Equation Solutions}

\subsection{Large Layer Spacing Limit}

Equations (\ref{flow1}) and (\ref{flow2}) can be combined to give a single nonlinear first-order differential equation
\begin{eqnarray}
\gamma\lambda_-\frac{d V}{d\lambda_-} = - \lambda_- + \alpha_+ - V^2.
\end{eqnarray}
This equation can be solved analytically with the result,
\begin{equation}
V(\lambda_-) = \frac{\sqrt{\gamma \lambda_-} \left(c J_{-\frac{2 \sqrt{\alpha_+}}{\gamma}-1}\left(\frac{2 \sqrt{\lambda_-}}{\sqrt{\gamma}}\right)-c J_{1-\frac{2 \sqrt{\alpha_+}}{\gamma}}\left(\frac{2 \sqrt{\lambda_-}}{\sqrt{\gamma}}\right)+J_{\frac{2 \sqrt{\alpha_+}}{\gamma}-1}\left(\frac{2 \sqrt{\lambda_-}}{\sqrt{\gamma}}\right)-J_{\frac{2 \sqrt{\alpha_+}}{\gamma}+1}\left(\frac{2 \sqrt{\lambda_-}}{\sqrt{\gamma}}\right)\right)}{2 \left(c J_{-\frac{2 \sqrt{\alpha_+}}{\gamma}}\left(\frac{2 \sqrt{\lambda_-}}{\sqrt{\gamma}}\right)+J_{\frac{2 \sqrt{\alpha_+}}{\gamma}}\left(\frac{2 \sqrt{\lambda_-}}{\sqrt{\gamma}}\right)\right)}
\label{general}
\end{equation}
where $c$ is an integration constant to be fixed by the UV boundary condition $V(\lambda_ = 0) = -\alpha_-+V_0$.  The value of $\lambda_-$ for which $V$ then goes to $-\infty$ then corresponds to a value of $\omega$ equal to the zero frequency gap $\Delta(0)$.

It is apparent from the flow diagrams (Fig.~\ref{flow}) that, for the gauge pairing case, in the limit of large layer spacing with $\alpha_ \ll 1$ with $V_0 > -\sqrt{\alpha_+}$, the solutions flow to the scaling solution flowing from the point $V = \sqrt{\alpha_+}$, $\lambda_= 0$.  This solution corresponds to (\ref{general}) with $c=0$,
\begin{equation}
V_{lls}(\lambda_-) = \frac{\sqrt{\gamma \lambda_-} \left(J_{\frac{2 \sqrt{\alpha_+}}{\gamma}-1}\left(\frac{2 \sqrt{\lambda_-}}{\sqrt{\gamma}}\right)-J_{\frac{2 \sqrt{\alpha_+}}{\gamma}+1}\left(\frac{2 \sqrt{\lambda_-}}{\sqrt{\gamma}}\right)\right)}{2 \left(J_{\frac{2 \sqrt{\alpha_+}}{\gamma}}\left(\frac{2 \sqrt{\lambda_-}}{\sqrt{\gamma}}\right)\right)}
\end{equation}

The energy gap $\Delta(0)$ is then determined by finding where $V_{lls} \rightarrow -\infty$ as $\omega$ is decreased from $\omega_0$ to $0$, which occurs when the Bessel function in the denominator has its first zero, i.e., when,
\begin{eqnarray}
2\sqrt{\frac{\lambda_-}{\gamma}} = J_{2\sqrt{\alpha_+}/\gamma,1}
\end{eqnarray}

The result for $\Delta(0)$ can then be expressed as follows
\begin{eqnarray}
\Delta(0) = \left(\frac{j_{0,1}}{j_{2\sqrt{\alpha_+}/\gamma,1}}\right)^{2/\gamma} \Delta(0;0)
\label{gap}
\end{eqnarray}
where
\begin{eqnarray}
\Delta_{\alpha_+=0}(0) = \frac{1}{j_{0,1}^{2/\gamma}} \left(\frac{4}{\gamma}\right)^{1/\gamma} = \left(\frac{C}{\gamma}\right)^{1/\gamma} \alpha_-^{1/\gamma} \omega_0.
\end{eqnarray}
where
\begin{eqnarray}
C = \frac{4}{j_{0,1}^2} = 0.69166\cdots
\end{eqnarray}

The multiplicative factor which suppresses the gap compared to its $\alpha_+ = 0$ value is  
\begin{eqnarray}
\frac{\Delta(0;\alpha_+)}{\Delta(0;0)} = 
\left(\frac{j_{0,1}}{j_{2\sqrt{\alpha_+}/\gamma,1}}\right)^{2/\gamma} \label{sfeq}
\end{eqnarray}
To get a simpler expression that captures the relevant behavior of this function, consider the limit where
$\gamma\rightarrow 0$ while $\sqrt{\alpha_+}/\gamma^2$ is held fixed.  In this limit, we can Taylor expand the quantity in parenthesis in (\ref{sfeq}), to find
\begin{eqnarray}
\frac{j_{0,1}}{j_{2\sqrt{\alpha_+}/\gamma},1} = 
1 - D \frac{\sqrt{\alpha_+}}{\gamma^{2}}\frac{\gamma}{2}  + \cdots
\end{eqnarray}
where
\begin{eqnarray}
D = 4 \frac{d}{dx} \ln j_{x,0} \Bigr|_{x=0} = 2.566 \cdots
\end{eqnarray}
We then have, (again, keeping $\sqrt{\alpha_+}/\gamma^2$ fixed),
\begin{eqnarray}
\lim_{\gamma \rightarrow 0 } \frac{\Delta(0;\alpha_+)}{\Delta(0;0)} &=& 
\lim_{\gamma \rightarrow 0} \left(\frac{j_{0,1}}{j_{2\sqrt{\alpha_+}/\gamma,1}}\right)^{2/\gamma}
= \lim_{\gamma\rightarrow 0} \left(1-D \frac{\sqrt{\alpha_+}}{\gamma^2} \frac{\gamma}{2}\right)^{2/\gamma} 
= \exp(-D \frac{\sqrt{\alpha_+}}{\gamma^2})
\end{eqnarray}
Finally, the frequency-dependent gap function $\Delta(\omega)$ can be obtained using (\ref{vtod}) with the overall scale set by $\Delta(0)$.

\subsection{Gauge Paring to BCS Crossover}

The crossover trajectory corresponds to the solution for which $V$ just touches the $V=0$ axis, without changing sign (and hence the energy gap doesn't change sign).  The value of $c$ for this case is easily obtained by setting $V(\lambda_- = \alpha_+/\gamma) = 0$, which fixes the trajectory to include the point in the flow diagram where the $\frac{dV}{dl} = 0$ curve (shown in black in Fig.~\ref{flow}) intersects the $V=0$ axis.  The result is
\begin{eqnarray}
c(\alpha_+,\gamma) =\frac{J_{\frac{2 \sqrt{\alpha_+}}{\gamma}+1}\left(\frac{2\sqrt{\alpha_+}}{\gamma}\right)-J_{\frac{2 \sqrt{\alpha_+}}{\gamma}-1}\left( \frac{2\sqrt{\alpha_+}}{\gamma}\right)}{J_{-\frac{2 \sqrt{\alpha_+}}{\gamma}-1}\left(\frac{2\sqrt{\alpha_+}}{\gamma}\right)-J_{1-\frac{2 \sqrt{\alpha_+}}{\gamma}}\left(\frac{2\sqrt{\alpha_+}}{\gamma}\right)}.
\end{eqnarray}
Plugging this value of $c$ back into the expression for $V$ gives $V(\omega)$ for this case which again can be used to determine the corresponding $\Delta(\omega)$ using (\ref{vtod}), which just touches the $\omega$ axis, as shown in Fig.~\ref{co}.

\twocolumngrid

\bibliography{references}

\end{document}